\documentclass[12pt,preprint]{aastex}
\usepackage{amssymb, amsmath}

\shorttitle{Partially Obscured QSOs} \shortauthors{Dong et al.}

\received{2003 January 28}
\begin{document}

\title{Partially obscured quasars in the Sloan Digital Sky Survey Early Data Release}
\author{Xiao-Bo Dong, Hong-Yan Zhou, Ting-Gui Wang, Jun-Xian Wang, Cheng Li, and You-Yuan Zhou}
\email{xbdong@mail.ustc.edu.cn}

\begin{abstract}
We have compiled a sample of 21 low redshift ($z \lesssim 0.3$),
luminous active galactic nuclei (AGN) with large Balmer decrements
(H$\alpha $/H$\beta > 7$) using the galaxy and QSO catalogs of the
Sloan Digital Sky Survey Early Data Release. Using this sample we
attempt to determine the fraction of quasars with large internal
absorption. We find that these AGN have Strong [O III] $\lambda
$5007 and broad H$\alpha $ emission, and that starlight dominates
the spectra in the blue band, suggesting that these objects are
heavily reddened. Their narrow emission line ratios are similar to
those of Seyfert 2 galaxies, yet the average [O III] $\lambda$5007
emission is $ \sim $10 times more luminous. Applying the empirical
relation between the optical continuum and the Balmer line
luminosity for blue quasars, we find that the intrinsic
luminosities of these 21 objects are in the range for quasars. We
propose that they are obscured, intermediate type quasars
analogous to type 1.8 and 1.9 Seyfert galaxies. The ratio of these
optically selected, intermediate type quasars to type 1 quasars
are found to be around 1, similar to that for local Seyfert
galaxies. Preliminary study indicates that most of these quasars
are hosted in early type galaxies.
\end{abstract}

\keywords{galaxies: active --- quasars: general --- galaxies: nuclei}

\affil{Center for Astrophysics, University of Science and
Technology of China, Hefei, 230026, China}

\section{Introduction}

The unified model for Seyfert galaxies is now well-supported (see
Antonucci 1993 for a review). The model postulates that Seyfert 1
and Seyfert 2 galaxies are physically the same, except that the
broad line region (BLR) in Seyfert 2 galaxies is obscured from our
line of sight by dust in the nucleus or in the host galaxy, and
that quasars are luminous versions of Seyfert galaxies.
Obscuration by dust should also play an important role in quasars.
However, the evidence supporting such unification for quasars is
substantially weaker than for Seyfert galaxies. Whether or not the
difference between Seyfert galaxies and QSOs is simply a matter of
luminosity is still a hotly debated issue, for instance some
researchers argue that there may not even exist a dusty torus in
QSOs, or, if one exists, its opening angle is luminosity dependent
(e.g., Lawrence 1991; Simpson 2003).

On the other hand, the hard X-ray spectrum of the cosmic X-ray
background (XRB) seems to require a large number of obscured AGN
(Seyfert galaxies and quasars). During the past few years, deep
X-ray surveys (ROSAT, Chandra and XMM-Newton) have resolved more
than 80\% of the 2-10 keV X-ray background into discrete sources
(e.g., Mushotzky et al. 2000; Hasinger et al. 2001; Rosati et al.
2002). Follow-up spectroscopic observations show that the sources
producing the bulk of the XRB are a mixture of obscured and
unobscured AGN, as predicted by the XRB population synthesis
models (Gilli et al. 2001 and references therein). However, the
ratio between obscured and unobscured quasars estimated according
to the black hole mass function in the local universe is much
lower than required by the XRB population synthesis models
(Marconi \& Salvati 2002; Yu \& Tremaine 2002).

The luminous analogs of Seyfert 2 galaxies predicted by the
unification model are hard to select and identify for the
following reasons: First, they are much rarer than Seyfert 2
galaxies, thus a large area deep survey is required. Second, it is
hard to obtain their intrinsic luminosities because of the heavy
obscuration, thus it is difficult to distinguish them from Seyfert
2 galaxies. Soft X-ray surveys, such as ROSAT, can not find such
objects effectively because the strong obscuration blocks soft
X-rays. Radio surveys can be affected by free-free opacity
associated with the obscuration, and moreover, only $\lesssim
10\%$ AGN are radio-loud (Krolik, 1999). Hard X-rays should be
relatively unaffected by extinction, but a large-scale, deep
survey has not yet been carried out. Though every now and then
there are reports of the discovery of type 2 QSOs\footnote{Some of
these objects may not follow the strict definition of the
\textquoteleft type 2\textquoteright\ AGN, since they sometimes
reveal weak broad H$\alpha $ lines in deeper, high S/N ratio
spectra (e.g.,1E0449.4--1823, Boyle et al. 1998; AX J08494+4454,
Akiyama et al. 2002). They are, however, heavily obscured AGN.}
(e.g., Norman et al. 2002; Della Ceca et al. 2003), the only
systematic search for such objects covering a large area is the
2MASS red AGN survey (Cutri et al. 2001).

The Sloan Digital Sky Survey (SDSS: York et al. 2000) is both an
imaging and spectroscopic survey. Its large sky coverage
($\thicksim 10^{4}$ deg$^{2}$) makes it ideal for selecting rare
objects such as heavily obscured quasars. In this paper we present
a sample of 21 intermediate type quasars selected from the SDSS
EDR (Early Data Release). They are selected to have strong, broad
H$\alpha $ emission lines, with a large Balmer decrement for the
broad component (H$\alpha $/H$\beta > 7$), suggesting strong,
partial obscuration/reddening. Their intrinsic luminosities are
estimated from their reddening-corrected broad H$\alpha $
emission. Their nearness ($z\lesssim $ 0.3, with a median value of
$z = 0.163$) makes them relatively easy for follow-up
observations. These objects are important for three reasons:
First, they can be used to study the nature of the obscuring
material in quasars. Second, the relative number of such objects
may provide us with some clues to the abundance of type 2 quasars,
provided they are obscured by the same mechanism. Finally, they
can be used to unravel the relationship between the nuclear
activity and the character of the host galaxy. Usually, the study
of QSO host galaxies is very difficult, especially for luminous
QSOs, because the light from the AGN itself overwhelms that from
the host galaxy. Thus, further work on the intermediate type QSOs
discussed in this paper, where the nuclear emission is strongly
attenuated and the host galaxy has prominent observed features,
may be able to yield less ambiguous results on the morphology,
stellar velocity dispersion, and star formation history in these
host galaxies.

This paper is organized as follows: In Section 2 we present the
details of our analysis of the SDSS spectra, and describe our
selection criteria for intermediate type QSOs. Section 3 presents
the main properties of our sample, as well as discussions of some
interesting individual objects. Section 4 is a short summary.
Through out the paper, we assume a cosmology with H$_{0}$=70
km~s$^{-1}$~Mpc$^{-1}$, $\Omega_{M} $=0.3 and
$\Omega_{\Lambda}$=0.7.

\section{Sample construction and Data Analysis}

\subsection{Construction of the parent sample}

The parent sample of our partially obscured QSOs is the low
redshift AGN (QSOs and Seyfert galaxies) with reliable broad
H$\alpha $ line selected from the spectral dataset of the SDSS
EDR. We construct the parent sample in four stages.

First, we pick up all the objects classified by the SDSS
spectroscopic pipeline as galaxies (SPEC\_GALAXY, specClass=2) or
quasars (SPEC\_QSO, specClass=3) with $z\lesssim 0.3$. This yields
38223 objects in total (37657 as SPEC\_GALAXY, 566 as SPEC\_QSO).
We limit $z\lesssim 0.3$ so that the centroid of H$\alpha $ lies
at $\lesssim $ 8500 \AA, allowing us to reliably measure the
H$\alpha $ flux. In this first step we do not limit the S/N ratio,
because some objects may have a reliable broad H$\alpha $ line yet
a rather weak continuum with a very low S/N ratio. For any object
with multiple spectroscopic observations, we retain only the
spectrum with highest S/N ratio.

Second, the 566 SPEC\_QSO are visually examined to remove 1)
objects with too many bad pixels or too low S/N ratio in the
H$\alpha $ region, and 2) objects obviously without broad emission
lines. This stage cuts the sample to 459 SPEC\_QSO. Using these we
build a sub-sample of normal type 1 AGN (QSOs and luminous Seyfert
1.0 galaxies, hereafter called blue AGN) to use as a reference in
the next stages. These blue AGN are selected according to the
following criteria: 1) $ D_{4000} \lesssim 1$ (the definition of
the 4000 \AA\ break index $D_{4000}$ introduced by Balogh et al.
(1999) is adopted here, using bands 3850-3890 \AA\ and 4000-4100
\AA ) , 2) blue color ($u^{\prime }-g^{\prime }\lesssim $ 0.6
corresponding to the conventional criterion of $U-B \lesssim $
-0.4). These criteria produce a sub-sample of 94 objects. For
these objects, we fit the continua using a single power-law or
broken power-law. For some objects with apparent Fe II multiplets,
we fit the continua employing the empirical optical Fe II template
of Boroson \& Green (1992, hereafter BG92) broadened by convolving
a Gaussian of various widths. The continuum (and FeII multiplets
if present) was then subtracted and the remaining emission lines
were fitted with an appropriate line profile (the details of the
emission line fitting are described in \S\ 2.3). These emission
line fits yield the equivalent widths (EW) of the broad H$\alpha$
component of the 94 objects. These range from $134.4$ to $726.2$
\AA\ with a mean value of $335.4$. These EW will be used later to
roughly evaluate the relative contribution of the stellar and
nuclear components.

In the third stage, nuclear continuum and/or starlight subtraction
is performed for the remaining 365 SPEC\_QSO and the 37657
SPEC\_GALAXY. For 41 SPEC\_QSO without prominent stellar features,
power law continuum (and Fe II templates if necessary) were fitted
in the same way as for the blue AGN. For the other SPEC\_QSO,
where the stellar contribution cannot be neglected, nuclear
continuum and stellar component are fitted simultaneously (see \S\
2.2).

For the 37657 SPEC\_GALAXY, we initially perform starlight
subtraction by fitting the emission line free regions of the
spectrum with nine synthesized templates, which will be discussed
in detail in \S\ 2.2. Then the emission lines of these 37657
objects are measured assuming Gaussian profiles. This is a
preparatory fitting for subsequent reductions. More than one
thousand candidates were selected using the criterion that the
broad component of H$\alpha $ (Full Width at Half Maximum,
FWHM$>1000$ km~s$^{-1}$) is significant ($>5 \sigma$). Then we
visually examined these candidates to exclude the objects with a
false broad component of H$\alpha$ caused by poor starlight
subtraction. After these culls, the number of the candidates is
reduced to 867. However, for some of them, the nuclear
contribution is not negligible. Those 126 objects with
$EW(H\alpha)>30~\AA$ ($\sim $0.2 times the minimum EW(H$\alpha$)
of blue AGN and $\sim $0.1 times the mean value) are picked up for
more precise starlight and nuclear continuum subtraction, as
described in \S\ 2.2.

The main task of the fourth stage in our analysis is to fit the
emission lines precisely using several schemes (see  \S\ 2.3) for
all the selected 1232 objects (459 SPEC\_QSO and 867
SPEC\_GALAXY). Based on the emission line parameters, we select
out 583 objects (including the 94 blue AGN) as our parent sample.
The criterion is: $f_{Ha}^{broad}>5$ $\sigma _{Ha,broad}^{total}$,
where $\sigma _{Ha,broad}^{total}$ is the total flux error of the
broad H$\alpha $ component synthesized from two parts: 1) $\sigma
_{Ha,broad}^{fit}$, the error given by the spectral line fitting
procedure and 2) $\sigma _{Ha,broad}^{sub}$, the error introduced
by template subtraction because of the mismatch between the
object's spectrum and the model spectrum. The latter will be
discussed in \S\ 2.2.

\subsection{starlight/continuum subtraction}

Accurate starlight subtraction for the measurement of nuclear
emission lines is important. Usually it is implemented by fitting
the non-emission-line regions of the spectrum with a model
spectrum composed of galaxy templates which are derived from
either observed star spectra (e.g., Kauffmann et al. 2003) or
observed galaxy spectra (e.g., Ho et al. 1997a). In this work we
develop a \textquotedblleft two-step\textquotedblright\ method to
build the galaxy templates described in detail in Li et al.
(2004). Here we provide a brief summary of this approach.

First, Principal Component Analysis (PCA) is applied to an
observed stellar library newly available (STELIB, Le Borgne 2003).
The resultant star eigenspectra broadened to various velocity
dispersions are then used to fit the high quality spectra of more
than one thousand galaxies selected from the SDSS spectral dataset
according to 6 color-color diagrams using 18 synthesized
magnitudes, which constitute a uniform library with a full
spectral type coverage. Based on the fitting result, we construct
model spectra of the 1016 representative galaxies with a zero
velocity dispersion. Again PCA is applied to these model spectra
and the galaxy eigenspectra with a zero velocity dispersion are
obtained. After test with SDSS galaxies using an F-test, the first
9 galaxy eigenspectra are picked up as our galaxy templates. Our
galaxy templates have three advantages in addition to the merits
of the PCA technique itself: 1) they do not include any light from
the AGN, since they are derived from the model spectra generated
from pure star eigenspectra; 2) they represent the most prominent
features of the SDSS galaxies since they are based on
representative galaxy spectra from the SDSS; 3) they can be used
to fit the stellar velocity dispersion of the galaxies since they
themselves are not velocity-broadened.

When used to fit the host galaxy component of the objects, these 9 templates
are broadened by convolving with a Gaussian to match the stellar velocity
dispersion of the host galaxy. The host galaxy component is thus modelled as
the linear combination of the 9 best fitting broadened spectra.

In our sample, many objects appear to have both a host galaxy
component and a nuclear component. In such cases, we fit the
(almost) emission-line-free regions of the spectrum with the 9
broadened template spectra + a reddened power-law with $E_{B-V}$
varying between 0 and 2.5. The latter represents the nuclear
component. Here the power-law is $f(\lambda )=C\lambda ^{-1.7}$,
taken as the unreddened nuclear continuum according to Francis
(1996). Some objects have visible Fe II emissions which must be
also removed before spectral line fitting. For those objects, we
first subtract the starlight/continuum component without taking
account of the Fe II emission and fit the spectral lines. Then we
broaden BG92's optical Fe II template to the width of the broad
H$\beta $, and fit the pseudo-continuum (wavelength ranges of
prominent emission lines other than FeII multiplets were masked
out) again with the 9 broadened template spectra + the reddened
power law + the broadened BG92's template.

In order to estimate the mismatch of this subtraction, we fit
$\sim 1000$ high quality spectra of absorption line galaxies and
obtain the standard deviation of the relative error of the
subtraction $\sigma _{r}$ (=0.027). Using this we estimate the
error of the flux of a broad line caused by the subtraction as the
flux of the underlying continuum within the region of 4 FWHMs of
the line multiplied by $\sigma _{r}$.

\subsection{Emission line fitting}

We developed a systematic code to fit the emission lines. With
this code, we can fit each spectrum using several different
fitting schemes. For example, for narrow H$\alpha $, we can tie
its width to that of another line, such as [S II] $\lambda $6731
while fitting the spectra with a extremely broad H$\alpha $
component; or we can just fix it to a designated value, such as
the fitted width of [O III] $\lambda $5007 when [O III] $\lambda
$5007 is the only reliable narrow line. The results of different
fitting schemes are compared with each other and the one with
minimum $\chi^2$ is chosen. In this work, the primary goal of the
line fitting is to obtain accurate measurements of broad H$\alpha$
and broad H$\beta $, with an emphasis on the former. To achieve
this goal, we must accurately separate the Balmer lines from the
nearby narrow lines. In general, our fitting strategies are as
follows:

1. Initially, lines are fitted with Gaussians, one Gaussian for each narrow
or broad component of the lines. For example, four Gaussians are used to fit
the H$\alpha $ + [N II] lines. The flux ratio of the [N II] doublet $\lambda $6583/$%
\lambda $6548 is fixed to the theoretical value of 2.96, and their
profiles are assumed to be identical. Likewise the [O III] doublet
is similarly constrained. Usually we also require the profiles of
the narrow H$\alpha $ and [N II] to be the same.

2. If the FWHM of the broad H$\alpha $ line is less than 3000
km~s$^{-1}$, we also try a Lorentzian profile to fit the broad
lines.

3. If [O III] or [S II] doublets show a complex profile (e.g.,
with faint, extended wings), double Gaussians are employed to fit
each narrow line.

4. If the broad and narrow H$\alpha $ components can not be
separated by the fit, the narrow lines in the Ha + [N II] region
will be given the same profile of the narrow H$\beta $ line if
available, or [S II], or [O III] in this order.

5. Likewise, if the broad and narrow H$\beta $ components can not
be separated by the fit, the narrow H$\beta $ line will be given
the same profile of the narrow H$\alpha $, or [O III], or [S II]
in this order.

6. If the broad H$\beta $ line is weak or unavailable, we
constrain its profile to be that of broad H$\alpha $ line.

Following the above steps, we have measured the lines of most
objects. But exceptions always exist since, as we know, the
profiles of emission lines may be formed in quite different
regions. For example, in several spectra the broad H$\alpha $
component shows two or more velocity peaks, while in some others
it even has an asymmetric or irregular shape. In these cases we
adopt multiple Gaussians to constrain the fit if the data quality
is high enough. (cf., 2 Gaussians for the broad H$\alpha$
component of SDSS J101405.89+000620.3 as plotted in Figure 1). In
fact, all the spectra are fitted interactively and then inspected
visually. The fitting of the H$\alpha $ + [N II] region for all
the 21 partially obscured QSOs are displayed in Figure 1 (the
right panel).

\bigskip

\subsection{The sample of intermediate type quasars}

It is well established that the Balmer line luminosity is tightly
correlated with the nuclear continuum luminosity for both quasars
and Seyfert galaxies (Yee, 1980; Shuder, 1981; Ho \& Peng, 2001).
We plot in Figure 2 the H$\alpha $ luminosity versus the absolute magnitude
for our blue AGN. A linear fit shows
\begin{equation}
\log L_{H\alpha }=-(0.340\pm 0.018)M_{g}+(35.17\pm
0.40)~(\mathrm{erg\;s}^{-1})
\end{equation}
which is perfectly consistent with the result of Ho \& Peng (2001) who got
\begin{equation}
\log L_{H\beta }=-(0.340\pm 0.012)M_{B}^{nuc}+(35.11\pm 0.25)
\end{equation}
for a sample of PG quasars and Seyfert galaxies. According to
equation (1), an H$\alpha $ luminosity of log~$L_{H\alpha }
(\mathrm{erg~s}^{-1})\simeq \allowbreak 42.82$ mag gives
$M_{g}=-22.5$ mag as the quasar selection limit in this paper.

Figure 3 shows the distribution of broad H$\alpha $/H$\beta $ for the 94
blue AGN we
selected. It can be seen that the scattering of the H$\alpha
$/H$\beta $ ratios is rather small, with an average H$\alpha
$/H$\beta $ of 2.97 $\pm$ 0.36.
Assuming that all the objects in the parent sample have the same
intrinsic Balmer decrement, H$\alpha $/H$\beta $=2.97,
we calculate the internal extinction as well as the
intrinsic luminosity of H$\alpha$ for the reddened objects.
The Galactic extinction curve in Fitzpatrick (1999) with Rv=3.1 is utilized and
the intrinsic luminosity of the broad H$\alpha $ component is estimated as:
\begin{equation}
\log L_{H\alpha }^{int}=\log L_{H\alpha }^{obs}+1.87(\log (H\alpha /H\beta)-\log (2.97))
\end{equation}

Figure 4 shows the observed luminosity of the broad H$\alpha $
component versus the H$\alpha $/H$\beta $ ratio for all the 583
objects in the parent sample. The tilted line corresponds to our
quasar selection limit, i.e., $M_{g,Est}^{nuc}\simeq-22.5$ mag or
$logL_{H\alpha }^{int}\simeq 42.82$ erg s$^{-1}$. Out of the 80
objects above the tilted line, we compile a partially obscured QSO
sample of 21 objects with H$\alpha $/H$\beta $ $>$ 7 (listed in
Table 1). The H$\alpha $/H$\beta $ $>$ 7 criterion is set to pick
up those objects with relatively severe obscuration, in analogy
with type 1.8/1.9 Seyfert galaxies. By this criterion, the sample
satisfies the requirement of the intrinsic nuclear luminosity 10
times greater than the observed one or E(B-V) $\gtrsim$ 0.75,
which is reddened more severely than the dust reddened quasars
(E(B-V) $\lesssim$ 0.5) compiled by Richards et al. (2003) from
the SDSS QSO catalog. The SDSS spectra of the 21 objects are shown
in Figure 1 (the left panel).

\clearpage \tablenum{1}
\begin{deluxetable}{llrrrrrl}
\tablecaption{Parameters of the 21 partially obscured quasars \label{tbl-1}}
\tablewidth{0pt}
\tablehead{
\colhead{Object} & \colhead{$z$} & \colhead{Flux} & \colhead{Flux\tablenotemark{a}}
& \colhead{Flux} & \colhead{FWHM} & \colhead{M$_{g}^{Est}$} & \colhead{Host} \\
\colhead{} & \colhead {} & \colhead{[O III] $\lambda $5007} & \colhead{H$\beta $, broad}
& \colhead{H$\alpha $, broad} & \colhead{H$\alpha $, broad} & \colhead{} & \colhead{} \\
\colhead{\tablenotemark{b} (1)} & \colhead{(2)} & \colhead{(3)} & \colhead{(4)} & \colhead{(5)} & \colhead{(6)} & \colhead{(7)} & \colhead{(8)}
}
\startdata
003511.48$-$004918.1 & .1859 & 206$\pm $6 & 254$\pm $17 & 1847$\pm $31 & 4298 & -23.2 & E\tablenotemark{d}       \\
003916.41$-$003232.8 & .1098 & 895$\pm $9 & 73$\pm $20 & 1081$\pm $21 & 4135 & -22.6 & E\tablenotemark{c}        \\
004527.06$+$004237.7 & .1095 & 2426$\pm $15 & $<$125 & 1968$\pm $34 & 3684 & $<$-23.6 & E\tablenotemark{c}       \\
015521.70$-$004150.0 & .2699 & 482$\pm $7 & 145$\pm $13 & 1366$\pm $39 & 4547 & -21.5 & ?                        \\
020036.60$+$004549.3 & .1672 & 278$\pm $4 & $<$57 & 765$\pm $25 & 10004 & $<$-23.2 & ?                           \\
023305.96$+$003856.4 & .2445 & 140$\pm $3 & 54$\pm $11 & 448$\pm $20 & 4555 & -22.5 & E\tablenotemark{d}         \\
030201.22$-$010117.9 & .1666 & 2492$\pm $15 & $<$91 & 1864$\pm $33 & 5176 & $<$-25.4 & ?                         \\
031142.02$-$005918.9 & .2815 & 123$\pm $4 & 47$\pm $12 & 481$\pm $30 & 5093 & -23.5 & E\tablenotemark{d}         \\
101405.89$+$000620.3 & .1411 & 522$\pm $8 & $<$340 & 5509$\pm $75 & 20854 & $<$-25.7 & E\tablenotemark{c}        \\
103352.59$+$004403.4 & .1312 & 2673$\pm $18 & 356$\pm $21 & 3558$\pm $41 & 5465 & -23.7 & S\tablenotemark{c}     \\
113021.41$+$005823.0 & .1324 & 1701$\pm $12 & 214$\pm $10 & 2552$\pm $15 & 6300 & -23.8 & E\tablenotemark{c}     \\
121026.40$-$002137.4 & .2915 & 421$\pm $4 & 98$\pm $10 & 724$\pm $28 & 4381 & -23.4 & E\tablenotemark{d}         \\
124321.77$+$001537.2 & .1426 & 729$\pm $8 & 54$\pm $12 & 760$\pm $16 & 2791 & -22.8 & E\tablenotemark{d}         \\
135717.59$+$002013.0 & .2834 & 239$\pm $3 & $<$40 & 419$\pm $33 & 5513 & $<$-23.5 & ?                            \\
141003.66$+$001250.2 & .1359 & 378$\pm $5 & 65$\pm $12 & 910$\pm $18 & 3728 & -22.9 & E\tablenotemark{d}         \\
143727.81$-$002343.5 & .1379 & 366$\pm $6 & 158$\pm $11 & 1639$\pm $17 & 3567 & -23.0 & E\tablenotemark{d}       \\
165641.98$+$632307.7 & .1610 & 385$\pm $7 & $<$13 & 385$\pm $12 & 2109 & $<$-21.0 & ?                            \\
170211.15$+$605848.1 & .1649 & 514$\pm $6 & 84$\pm $25 & 1036$\pm $31 & 10003 & -23.3 & ?                        \\
170601.87$+$601732.4 & .1303 & 249$\pm $4 & 69$\pm $17 & 1390$\pm $27 & 8076 & -21.2 & E\tablenotemark{d}        \\
171832.86$+$531304.7 & .1916 & 169$\pm $4 & 129$\pm $14 & 967$\pm $26 & 2637 & -22.5 & ?                         \\
233816.44$+$005029.7 & .1832 & 92$\pm $4 & 106$\pm $11 & 939$\pm $16 & 3943 & -22.7 & ?                          \\
\enddata

\tablenotetext{a}{Adopting the profile of broad H$\alpha $.}
\tablenotetext{b}{(1) hhmmss.ss$\pm $ddmmss.s; (3),(4) and (5) in
unit of 10$^{-17}$ erg~cm$^{-1}$~s$^{-1}$; (6) in unit of km~s$^{-1}$;
(7) estimated absolute $g^{\prime}$ magnitude based on the reddening-corrected luminosity
of broad H$_\alpha$ line; (8) Morphology of the host galaxy: E for early type (E, S0, Sa), S for late type. }
\tablenotetext{c}{Classified by visual inspection.}
\tablenotetext{d}{Classified according to likelihoods and concentration index.}

\end{deluxetable}
\clearpage

\section{Discussion}

\subsection{Testing the assumption of large, internal extinction}

We have tentatively classified 21 objects as intermediate type quasars
according to their high intrinsic nuclear luminosities estimated by equation
(3). The most remarkable property of these objects is their large H$\alpha $%
/H$\beta $ ratios. Our basic assumption is that their intrinsic H$\alpha $/H$%
\beta $ ratio is similar to that of the \textquotedblleft
normal\textquotedblright\ QSOs, i.e. 2.97, and that the large observed H$%
\alpha $/H$\beta $ ratio is due to relatively large internal extinction.
This assumption is consistent with the fact that the optical continuum of
almost all these objects is dominated by the stellar component of the host
galaxy (see Figure 1).

To test our assumption, we match the parent sample and 2MASS
catalogs and find 286 objects having counterparts within
$1\arcsec$ in the 2MASS Point Source Catalog (PSC), without any
extended \textquotedblleft contamination\textquotedblright and
with a reliable magnitude in the $K_{S}$ band. Of these objects,
the bigger the H$\alpha $/H$\beta $ ratio, generally the redder
the $u^{\prime} -K_{S}$ color is. All sources with H$\alpha
$/H$\beta $ ratio $>$ 7 have $u^{\prime} -K_{S}>4.5$ (Figure 5),
indicating their H$\alpha $/H$\beta $ ratios and $u^{\prime}
-K_{S}$ colors are consistent with our assumption. The absolute
$K_{S}$ magnitudes are well correlated with the luminosities of
the broad H$\alpha $ components of these objects (Figure 6). To
eliminate spurious correlations due to the redshift effect, we
calculate the partial correlation coefficient for a redshift, $z$,
as:
\begin{equation}
r_{xy,z}=\frac{r_{xy}-(r_{xz})(r_{yz})}{\sqrt{(1-r_{xz}^{2})}\times \sqrt{%
(1-r_{yz}^{2})}}
\end{equation}
where the $x$, $y$ refer to the H$\alpha $ luminosity and the
absolute $K_{S}$ magnitude, and $r_{xy}$,
$r_{xz}$ and $r_{yz}$ the correlation coefficients between $x$ and
$y$, $x$ and $z$, and $y$ and $z$. The partial correlation coefficient is
-0.625 with a chance probability of $\ll $0.0001 for
log$L_{\mathrm{H\alpha }}$ and $M_{K_{S}}$, and goes up in
absolute
value to -0.724 for log$L_{\mathrm{H\alpha }}^{int}$ and $%
M_{K_{S}}^{int} $ estimated according to our assumption. Again,
this favors our assumption of large internal extinction.

The extinction assumption is also substantiated by the relatively
high [O III] luminosity. In addition to H$\alpha $ luminosity,
there is observational evidence that [O III] luminosity can be
used as a rough estimate of the intrinsic power of an AGN, even in
type 2 objects (V{\' e}ron-Cetty \& V{\' e}ron 2000 and references
therein). Figure 7 plots the relation between [O III] luminosity
and the estimated nuclear absolute magnitude for the objects with
reliable [O III] $\lambda $5007 emission and reliable broad-line
Balmer decrements in the parent sample. The [O III] luminosities
are corrected for intrinsic extinction assuming an intrinsic
Balmer decrement equal to 3.1 for the narrow components of
H$\alpha $ and H$\beta $ (cf., Veilleux \& Osterbrock 1987), and
the intrinsic nuclear absolute magnitudes are estimated based on
the intrinsic luminosity of the broad H$\alpha $ component
according to equation (1) and equation (3). It is clear that the
two are correlated, though the scatter in this relation is much
larger than that for the correlation between the observed
continuum and the H$\alpha $ luminosity for the blue AGN (see
Figure 2). The derived partial correlation coefficient is -0.578
for log$L_{\mathrm{[OIII]}} $ and $M_{g}^{int}$ with a chance
probability of $\ll $0.001. Hence, the correlation indicated in
Figure 7 is not due to the redshift effect. The 87 QSOs/Seyfert 1s
in the BQS sample (Boroson \& Green, 1992) as well as all the
intermediate type quasars are both plotted for comparison. The [O
III] luminosity of these intermediate quasars are comparable to
that of luminous QSOs in the BQS sample and about 10 times more
luminous than that of Seyfert galaxies, suggesting that they very
possibly contain a quasar-like nucleus. However, from the
relatively greater scatter of the relation between
log$L_{\mathrm{[OIII]}}$ and $M_{g}^{int}$ for both the luminous
PG QSOs and our intermediate type QSOs, we may infer that the [O
III] luminosity is not a good calibrator for higher nuclear
luminosity.

The source of obscuration may be small scale dusty material near
the nucleus or a large scale dust lane in the host galaxy. For the
mean [O III] luminosity of our intermediate quasars, the narrow
line region (NLR) size is typically of kpc scale (Bennert et al.
2002). The kpc scale dust lane would also produce large reddening
in the NLR, while small scale dusty material would not.
To see which of these probabilities is most likely, we plot in Figure 8 the H$\alpha $/H$%
\beta $ ratio of the broad component versus that of the narrow
component for 141 objects with reliably measured, narrow
H$\alpha$, H$\beta $ lines, including 9 objects which are
intermediate type QSOs. First, it can be seen that our
intermediate type quasars, together with most of the other partly
obscured AGN, show no or moderate reddening in the NLR as
indicated by the narrow H$\alpha $/H$\beta $ ratio in the range
$2.0-5.5$, while some of them show heavy reddening in the BLR with
the broad H$\alpha $/H$\beta $ ratio $>$ 8 and up to $\sim $ 20.
Hence, the absorbing material could be the dusty torus similar to
that seen in Seyfert galaxies, or some other small scale obscuring
material in the nuclear region. However a patchy, large scale dust
lane cannot be ruled out. Second, for a rather larger number of
objects, however, both broad and narrow lines are comparably
reddened to some considerable extent (with the narrow and broad
H$\alpha $/H$\beta $ ratios $\sim $ 5), indicating that the
absorbing material might be a kpc scale dust lane or molecular
disk in the host galaxy, which can play a major role in producing
moderate, but not heavy, reddening.

\subsection{Emission lines and continua}

The broad H$\alpha $ FWHMs of the 21 objects scatter widely from
$\sim $2,100 km s$^{-1}$ (SDSS J165641.98+632307.7) to $>$20,000
km s$^{-1}$ (SDSS J101405.89+000620.3), with a median value of
$\sim $ 4500 km~s$^{-1}$.

In Figure 9 we plot [O III]$\lambda $5007/H$\beta $ versus [N
II]/H$\alpha $ and [S II]/H$\alpha$ respectively for the 21
quasars. Most of the objects are located in the AGN region of the
diagram (Veilleux \& Osterbrock, 1987) and the distribution is
similar to that of Seyfert 2 galaxies, indicating that these
objects are similar to normal Seyfert galaxies in the properties
of the NLR. Thus quasars, even bona fide type 2 quasars, should
also bear these properties. This result is consistent with the
recent HST observations of the NLR of bright radio-quiet PG QSOs
by Bennert et al. (2002), who also suggested that the NLR in
quasars is a scaled-up version of that of Seyfert galaxies.
However, there is one exception. SDSS J170601.87+601732.4 is
located at the border between AGN and LINERs, with [O II]$\lambda $3727/[O III]$%
\lambda $5007$>$1, yet [O I]$\lambda $6300/[O III]$\lambda
$5007$<$1/3. It has an unusual spectrum and in fact is the only
object in the parent sample that cannot be well fitted by our
galaxy templates. We will discuss this object in detail in \S\
3.5.

Our sample of intermediate type quasars can be considered to be
optically selected, since 19 of them are optical-color selected
for SDSS spectroscopic follow-up, and only 2 of them are selected
based on ROSAT detection. We find 6 of the 21 intermediate type
quasars have FIRST/NVSS\footnote{Faint Images of the Radio Sky at
Twenty Centimeters, Becker et al. 1995; NRAO VLA Sky Survey,
Condon et al. 1998.} counterparts within 5\arcsec, and the radio
detection rate is about 28$\pm $8\%. This ratio is slightly higher
than that of the 49 optically selected quasars (only 2 objects are
selected based on ROSAT detection) at z $<$ 0.3 in the SDSS Quasar
Catalog (Schneider et al. 2002), which is 14\%. Note an even
larger radio detection rate was reported by Richards et al.
(2001), who found that nearly half of the reddened quasars are
FIRST radio sources. There is only one radio-loud object in each
sample (Schneider et al.'s and ours). Moreover, each is the only
source resolved by FIRST.

\subsection{The fraction of partially obscured quasars}

In the same parent sample where we select our 21 intermediate type
quasars, we also pick up 44 type 1 QSOs. The selection criteria we
used are log~$L_{H\alpha }(\mathrm{erg~s}^{-1})\allowbreak \geq
42.82$ (i.e., $M_{g}\leq -22.5$), and small Balmer decrements:
$2.3\leq H\alpha ^{BC}/H\beta ^{BC}\leq 4.1$. The ratio of type 1
QSOs and the intermediate type-1s is 44:21$\approx $2.1:1. It
should be pointed out that the above two sub-samples may not be
complete, and the completeness of the samples is mainly affected
by two factors: 1) the quality of the spectra which influenced our
candidate selection and 2) the magnitude limit that SDSS used to
select targets for spectroscopic observation. The spectral quality
does not affect the selection of type 1 quasars, all of which have
high S/N spectra, however it is not negligible for the selection
of intermediate type quasars. In Figure 4 we can see that the main
incompleteness comes from the measurement of the broad $H\beta$
component. For many objects, due to the limited spectra quality,
we can only give upper limits to the broad $H\beta$ flux. Thus our
sample is conservative. A quick calculation shows that if we
decrease the upper limits by 50\%, 20 more sources would pass our
selection criteria. We also check the possible incompleteness due
to the requirement of at least a 5$\sigma$ detection of a broad
H$\alpha$ component. This potential selection effect is
negligible, as discussed below. We note that over half (13 of 21)
of the intermediate type QSOs are from the SDSS galaxy catalog
with a magnitude limit of 17.77, and that the typical S/N ratio at
this magnitude is 17.5. We can evaluate the limiting EW of the
broad H$\alpha $ line with this typical S/N ratio and the median
FWHM of the sample QSOs (4500 km~s$^{-1}$) corresponding to our
criterion for reliable broad H$\alpha $ line ($f>5\sigma $), which
turn out to be 8 \AA. Actually, the EWs of all the 21 objects are
greater than 30 \AA.

In order to correct the bias caused by the magnitude limit imposed
on the spectroscopic target selection, for each object $i$ we
calculate $z^i_{max}$ where the object would reach the magnitude
limit for SDSS spectroscopic follow-up. The corrected number of
sources is estimated as $\Sigma ~ V_{0.3}/V^i_{max}$, where
$V_{0.3}$ is the comoving volume within $z = 0.3$, and $V^i_{max}$
is that within the smaller of $z^i_{max}$ and 0.3. After the bias
correction for this imposed magnitude limit, the number of
intermediate type QSOs rises to 60.4 and the number of type 1 QSOs
remains untouched. The bias corrected ratio of type 1 QSOs to
intermediate type QSOs then becomes 44:60.4$\approx $1:1.4. To
compare this with Seyfert galaxies, we pick up 12 objects from our
intermediate type QSO sample which should be classified as type
1.8 or type 1.9 according to the quantitative classification
introduced by Winkler (1992), and the bias corrected number for
these type 1.8+1.9 objects is 39.1. Thus the ratio of type 1 QSOs
to the type 1.8+1.9 is 44:39.1 $\thickapprox$ 1:1, which is almost
the same as that of Seyfert galaxies (cf., Sy1.8+1.9:Sy1
$\thickapprox $ 1:1, Forster, 1999). But we must note that these
above ratios rely on the intrinsic H$\alpha$/H$\beta$ ratio and
the internal extinction curve we adopt for those intermediate type
AGN. For example, if the intrinsic H$\alpha$/H$\beta$ ratio is set
to 4.0, there are 13 intermediate type QSOs remaining and the bias
corrected ratio of type 1 QSOs to intermediate type QSOs changes
to 44:30.8$\approx $1.4:1.

This large fraction of obscured quasars have also been reported in
many other studies. Through studying the red quasars in the Parkes
Half-Jansky Flat-spectrum Sample, Webster et al. (1995) proposed
that 80\% of quasar population is dust-obscured and has been
missed by optical surveys, if the dust content around radio-quiet
quasars is as extensive as that around the radio-loud ones. Lacy
et al. (2002) also suggested that the red quasar population could
be very large through modelling the selection effects which turn
out to be effective at removing dust-reddened quasars from
magnitude-limited samples. Whiting et al. (2001), however, found
that $\sim$ 40\% of the Parkes sources show evidence for optical
synchrotron emission, which is responsible for the red color of
red radio-loud quasars in Webster et al.'s sample, as Serjeant \&
Rawlings (1996) argued. In that case, any missing population of
quasars would be much smaller. But a recent study by White et al.
(2003), based on a sample constructed by comparing the FIRST
survey with a $I$-band survey, strongly supports the hypothesis
that radio-loud and radio-intermediate quasars are dominated by a
previously undetected population of red, heavily dust-obscured
objects, although they do not know whether this kind of quasars
are common in the radio-quiet population. Our sample is almost
purely optically selected, with only one radio-loud object (see
\S\ 3.2), therefore we can safely conclude that dust-obscured
quasars are also common in the radio-quiet population. As stated
in \S3.1, the obscuring material should be on small scales in the
nucleus of the quasars.

Using the SDSS quasar catalog, Richards et al. (2003) found that
the redness of quasars whose colors cannot be fitted with a single
pow-law continuum is caused by dust extinction instead of
synchrotron emission, and moderately dust-obscured ( E(B-V)
$\lesssim$ 0.5) quasars account for 15\% of broad-line quasars. As
mentioned earlier, our sample is focused on heavily obscured
quasars (E(B-V) $\gtrsim$ 0.75), with over half of its members
from the SDSS galaxy catalog. Considering that Richards et al.'s
sample is restricted to the SDSS quasar catalog, our result is
marginally consistent with theirs.

\subsection{Properties of the host galaxies}

The host galaxy properties of quasars are generally difficult to
determine as the stellar component is usually overwhelmed by the
luminous quasars. High resolution ground and space observations
are usually the only way to detect the existence of the host
galaxy. Our sample of intermediate type quasar provides us with an
excellent opportunity to study their host galaxies, since their
nuclear emission is strongly attenuated by the obscuration. In
this subsection, we try to draw some straightforward conclusions
about the morphologies of the host galaxies. The study of star
forming history in the host galaxies of these objects is deferred
to another paper.

In Figure 10, we present the SDSS images.
For 5 bright objects with $r^{\prime}$ $<17$ mag, we classify the
type of host galaxy visually.
For the remainder, the criteria for ellipticals are:
\begin{itemize}
    \item{ concentration index $r_{90}/r_{50}\geq 2.5$ in $r^{\ast }$ (cf. Bernardi et al.
    2003);}
    \item{ the likelihood of the de Vaucouleurs profile fit is
    $>$ 0.5, and greater than the likelihood of the exponential
    model at least by 0.1.}
\end{itemize}
The criteria for spirals are:
\begin{itemize}
    \item concentration index $r_{90}/r_{50}<2.5$ in $r^{\ast }$;
    \item the likelihood of the exponential model fit is
    $>$ 0.5, and greater than the likelihood of the de Vaucouleurs profile fit at least by 0.1.
\end{itemize}
Following the above procedure, we classify 13 objects in total: 1
resides in a spiral, and 12 reside in ellipticals (listed in Table
1). The other 8 objects are left unclassified. This is consistent
with the result from HST observations that the host galaxies of
$\sim $2/3 QSOs are ellipticals (Hamilton et al. 2002).

Some objects show possible evidence for either interacting or
peculiar morphology. SDSS J170211.15+605848.1 has two companions,
with one at a very close projected distance ($\thicksim2\arcsec.5$
or less than 10 kpc physical distance). SDSS J170601.87+601732.4
may have companion galaxies also. SDSS J103352.59+004403.4 has a
compact core and a very extended (50 kpc in diameter) low surface
brightness disk viewed nearly edge on. SDSS J171832.86+531304.7
has a faint spiral structure, and could be a spiral galaxy.

\subsection{Notes on individual objects}

SDSS J101405.89+000620.3: The broad H$\alpha $ FWHM of this object
is extremely large (20078$\pm $419.8 km~s$^{-1}$). Its broad
H$\alpha $ profile has a flat-top, similar to the double peaked
line profile seen in some radio galaxies (e.g., Osterbrock, Koski,
\& Phillips 1976; Halpern \& Filippenko 1988; Halpern et al.
1996). It is a 2MASS red AGN (Cutri et al. 2001) with a red near
infrared color of J-K$_{s}$=2.012$\pm
$0.106 mag and is bright in the near infrared (J=15.240$\pm $0.061, H=14.311$%
\pm $0.057, and K$_{s}$=13.228$\pm $0.045).

Note 19 out of the 21 intermediate type quasars are in
the 2MASS Point Source Catalog, among these only 2 have J-K$_{s}$%
$>$ 2, 16 objects have 1.2 $< J-K_{s}$ $<$ 2, and 1 (SDSS
J170601.87+601732.4) has $J-K_{s}=1.041\pm 0.133$. Besides SDSS
J101405.89+000620.3, the other red object is SDSS
J103352.59+004403.4, with $J-K_{s}=2.073\pm 0.130$. This means
that most of the partially obscured quasars would be omitted by
the 2MASS red AGN survey ($J-K_{s}>$ 2). The relatively bluer
$J-K_{s}$ colors may be caused by the starlight from their host
galaxies.

SDSS J113021.41+005823.0: This object may be classified as a FR II
radio galaxy according to the usual classification criteria. Like
SDSS J101405.89+000620.3, its broad H$\alpha $ profile also has a
flat top. One interesting feature of this object is that it shows
rare X-shaped radio morphology with possible flows from the
central radio component to the wings. A detailed discussion of
this source was presented in Wang et al. (2003).

SDSS J135717.59+002013.0: This object may be taken as a heavily
reddened cousin of the famous \textquotedblleft post-starburst
quasar\textquotedblright\ UN J1025-0040 ($z=0.634$, Brotherton et
al. 1999) located at a lower redshift ($z=0.283$): its spectrum
displays prominent Balmer jump and high-order Balmer absorption
lines indicating a substantial, young stellar population. It is a
ROSAT X-Ray source, with a fairly high luminosity in the soft
X-ray band (L$_X$ = 7.21$\times $10$^{44}$ erg s$^{-1}$,
uncorrected for internal extinction). It has a rather flat ROSAT
X-ray spectrum with photon index $\Gamma$=1.12$_{-0.89}^{+0.51}$,
indicating a heavy internal absorption. It is also fairly luminous
in the near infrared, with $M_{K}=-25.9$ before internal
extinction correction.

SDSS J170601.87+601732.4: As mentioned in \S\ 3.2, its spectrum is
the only one in the parent sample that cannot be well fitted even
using 24 eigenspectra derived from the STELIB library. Its unusual
spectrum is probably not be due to contamination by a foreground
star, since the nearest star is 10$\arcsec$ far away. However, the
possibility of a foreground star sitting right on the galaxy can
not be completely ruled out. The poor spectral fit should not be
due to improper flux calibration or sky subtraction either, for
other spectra in the same plate look well-calibrated. In the light
of prominent absorption lines and absorption bands presented in
the spectrum, we propose that this spectrum is dominated by a very
old and high metallicity stellar population, which is beyond the
coverage of the STELIB library and thus cannot be modelled by our
galaxy templates which are based on the library. As noted in \S\
3.2, this object is LINER-like. Most LINERs are found to have
predominantly old stellar populations (Fernandes et al. 2004;
Gonz{\' a}lez Delgado et al. 2004), which reinforces our proposal.
We tend to believe that the broad H$\alpha$ component in this
object is real, but not formed by nearby absorption of neutral
metals and molecules in an old stellar population, because: 1) its
conspicuous and smooth line profile is different from that of the
other smaller bumps blueward of it, which are likely due to nearby
absorption, and 2) there is no strong absorption redward of the
H$\alpha$ emission line. Thus, SDSS J170601.87+601732.4 may be a
heavily reddened, luminous version of \textquotedblleft LINER
1s\textquotedblright\ (Ho et al. 1997b).

\section{Summary and Future Prospects}

We have compiled a sample of 21 objects with relatively strong broad-H$%
\alpha $ but very weak broad-H$\beta $ emission. Assuming that the
broad line is produced by photo-ionization process and that the
large Balmer decrement is due to internal reddening, the intrinsic
optical continuum luminosity of the nuclei, or the lower limit on
it, is estimated to have absolute magnitude M$_{g}\lesssim
$-22$^{m}$.5, implying that these are intermediate type quasars.
The ratio of these intermediate type quasars to type 1 quasars is
similar to that of intermediate type Seyfert galaxies to type 1
Seyfert galaxies, indicating that quasars may be somewhat similar
to Seyfert galaxies in structure. The narrow lines of these
intermediate type quasars are found to be similar to those of
Seyfert galaxies in line ratios, but more luminous. Our study has
revealed that these partially obscured quasars tend to be hosted
in early type galaxies more frequently than in late type galaxies.
thus, these partially obscured quasars are important for the
studies of the host galaxies related to morphology, stellar
velocity dispersion and stellar population, and also the relation
between the QSO phenomenon and the galactic evolution. A larger
sample with higher S/N ratio is being constructed based on the
observations in the SDSS Data Release 2.

Further multi-band observations are essential to reveal the nature
of these intermediate type quasars. Some individual objects in the
sample are also very interesting and deserve further research.
Hard X-ray observations, which can be done by XMM-Newton and/or
$Chandra$ (Wilkes et al. 2002), would enable us to study the
strong obscuration, and derive the intrinsic luminosity using an
independent approach. Optical polarization spectroscopy would also
help determine the geometry of scattering/obscuring matter inside
these objects, using an approach similar to that carried out for
infrared selected quasars (Smith et al. 2000).

\acknowledgements This work is supported by Chinese NSF grants
NSF-19925313 and NSF-10233030 and a key program of Chinese Science
and Technology ministry. This work has made use of the data
products of the SDSS.

Funding for the creation and distribution of the SDSS Archive has
been provided by the Alfred P. Sloan Foundation, the Participating
Institutions, the National Aeronautics and Space Administration,
the National Science Foundation, the U.S. Department of Energy,
the Japanese Monbukagakusho, and the Max Planck Society. The SDSS
Web site is http://www.sdss.org/. The SDSS is managed by the
Astrophysical Research Consortium (ARC) for the Participating
Institutions. The Participating Institutions are The University of
Chicago, Fermilab, the Institute for Advanced Study, the Japan
Participation Group, The Johns Hopkins University, Los Alamos
National Laboratory, the Max-Planck-Institute for Astronomy
(MPIA), the Max-Planck-Institute for Astrophysics (MPA), New
Mexico State University, Princeton University, the United States
Naval Observatory, and the University of Washington. This research
has made use of the NASA/ IPAC Infrared Science Archive, which is
operated by the Jet Propulsion Laboratory, California Institute of
Technology, under contract with the National Aeronautics and Space
Administration.

We wish to thank an anonymous referee for several very careful and
thorough reviews and valuable suggestions, and the editor who
spend plenty of time correcting the numerous English errors in our
paper before it can be accepted for publication.

\clearpage

\begin{figure}[tbp]
\label{fig-1a}\epsscale{0.8} \plotone{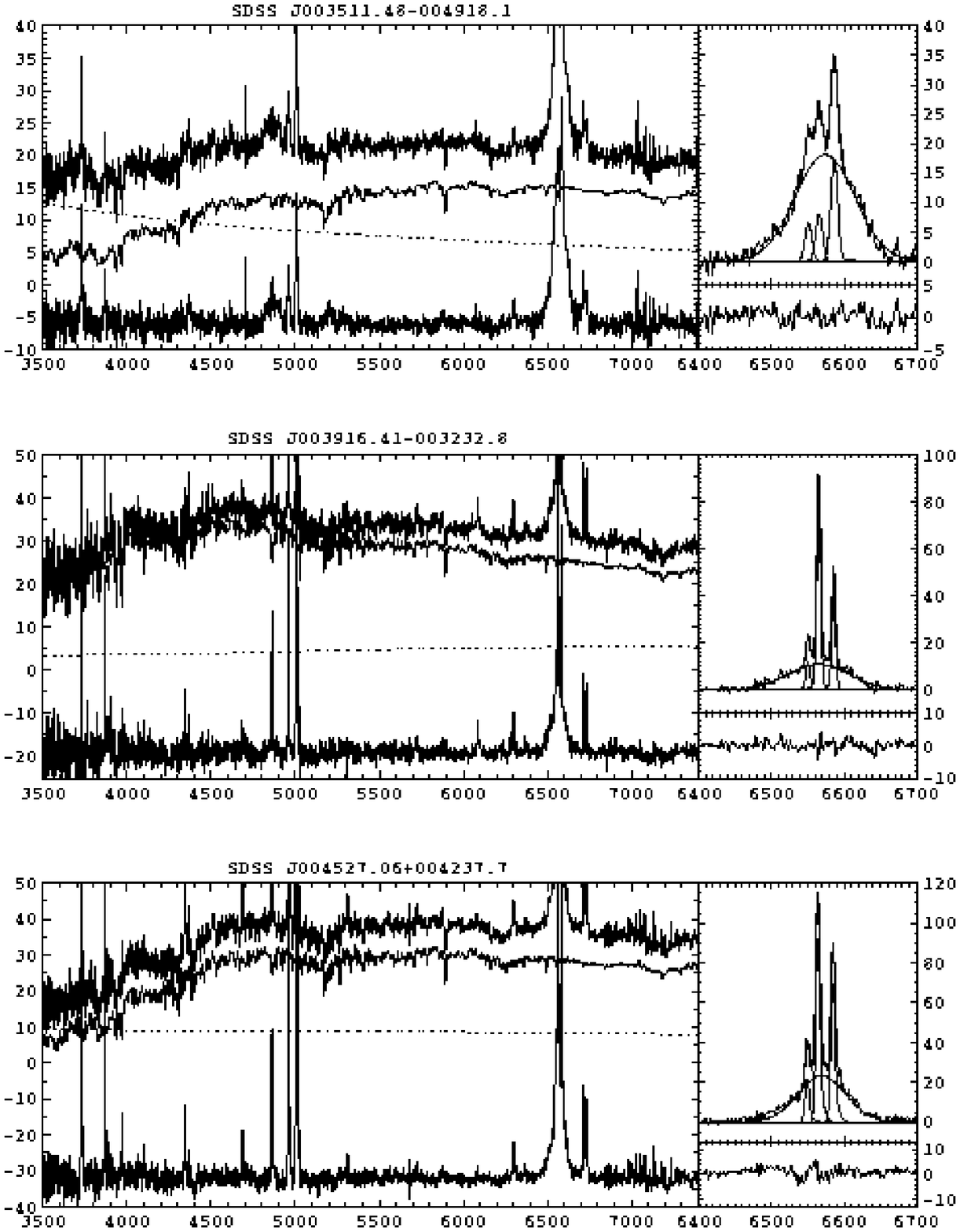} \caption{SDSS
spectra of the intermediate type quasars and the line fits. The
vertical axis is flux in unit of 10$^{-17}$
erg~s$^{-1}$~cm$^{-2}$~\AA $^{-1}$ and the horizontal is
wavelength in \AA . The left panel shows the procedure of
starlight/continuum subtraction: the original spectrum (top), the
stellar component (middle, solid line), the nuclear continuum
(middle, dotted line), and the starlight/continuum subtracted
residual (bottom). For clarity, the residual is offset downward by
an arbitrary constant. The right panel shows the process of line
parameter measurement. Top: original data and individual
components of the fit (thin curves), final fit (thick curve).
Bottom: the residual.}
\end{figure}

\figurenum{1}
\begin{figure}[tbp]
\epsscale{1} \plotone{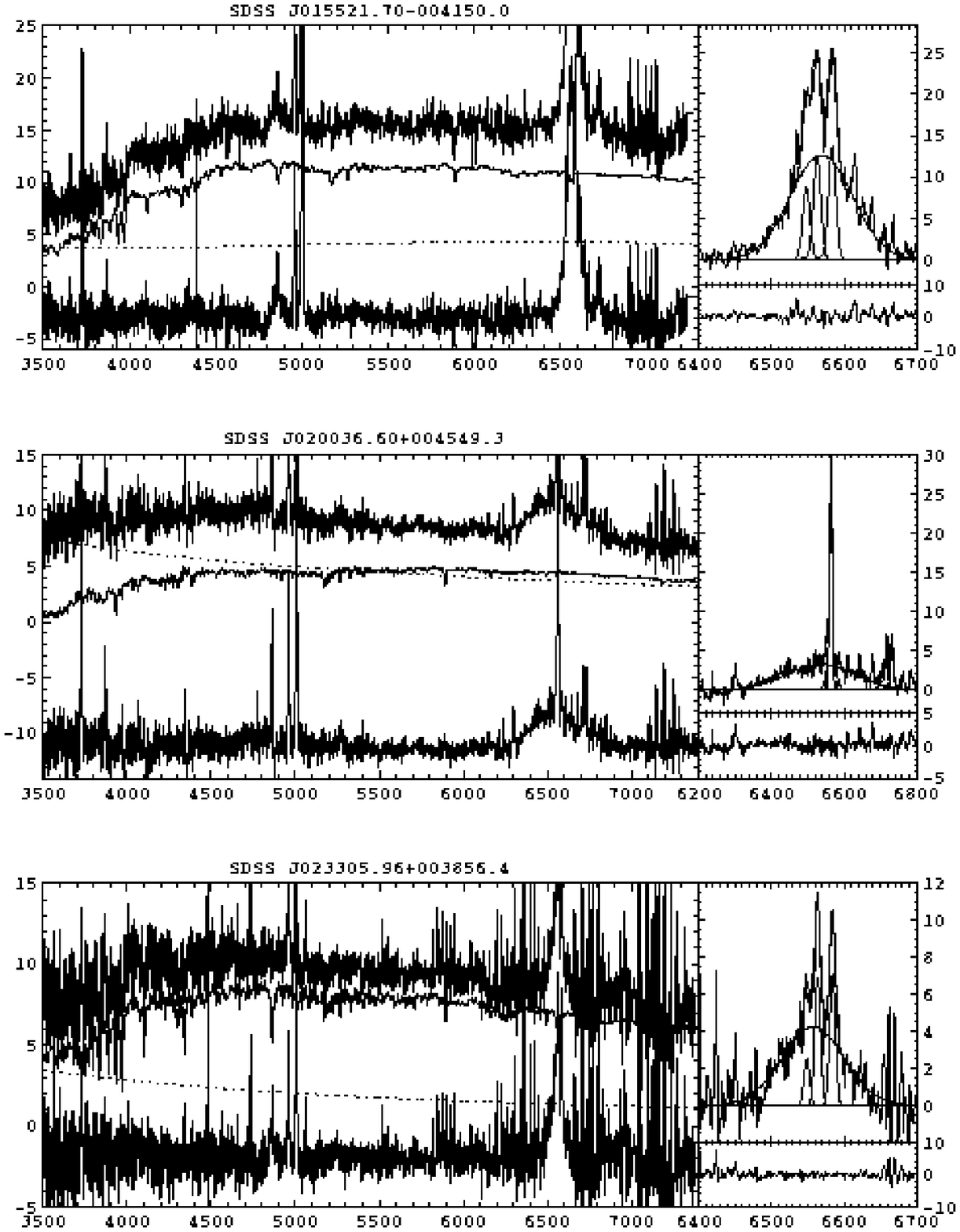} \caption{Continued...}
\end{figure}

\figurenum{1}
\begin{figure}[tbp]
\epsscale{1} \plotone{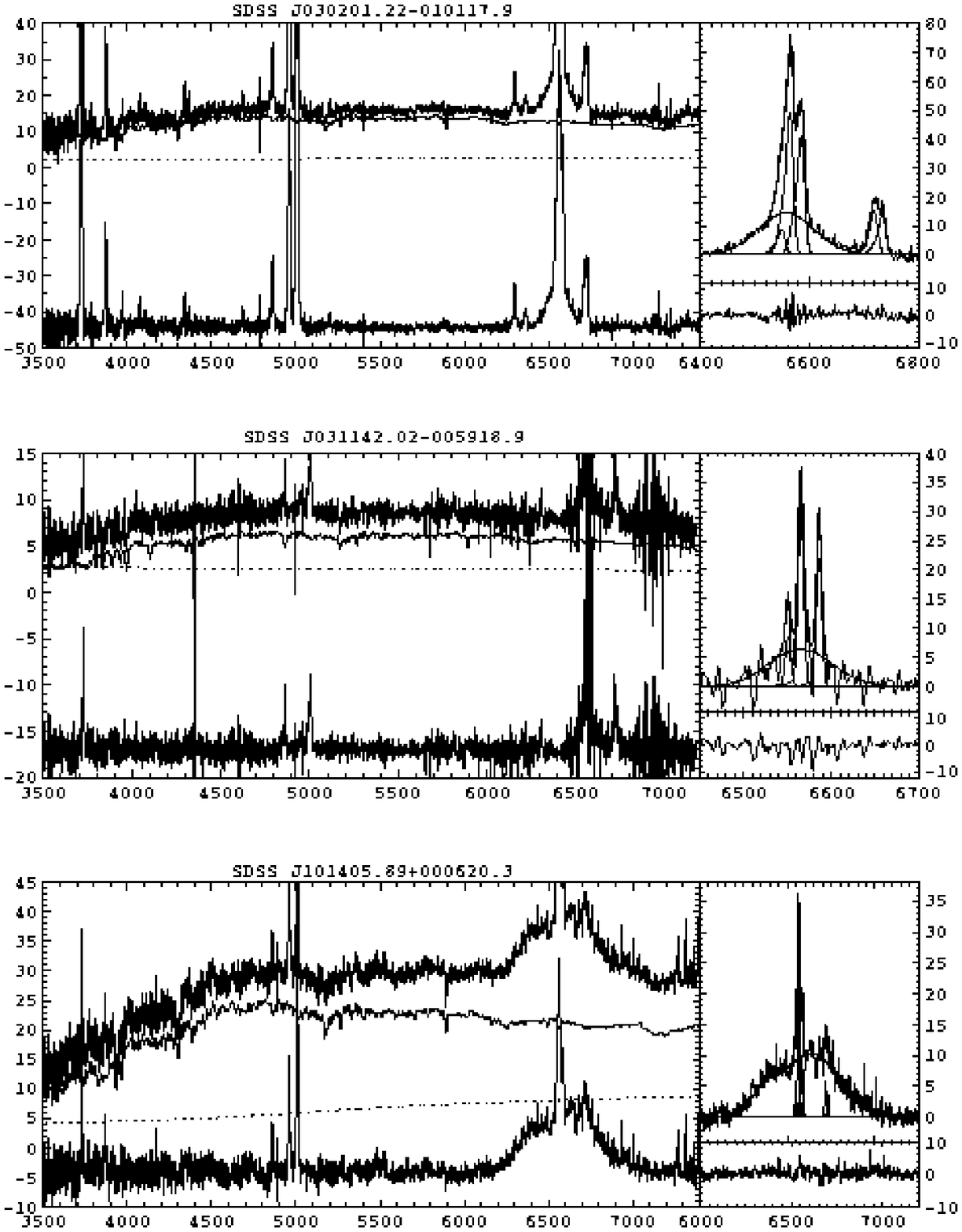} \caption{Continued...}
\end{figure}

\figurenum{1}
\begin{figure}[tbp]
\epsscale{1} \plotone{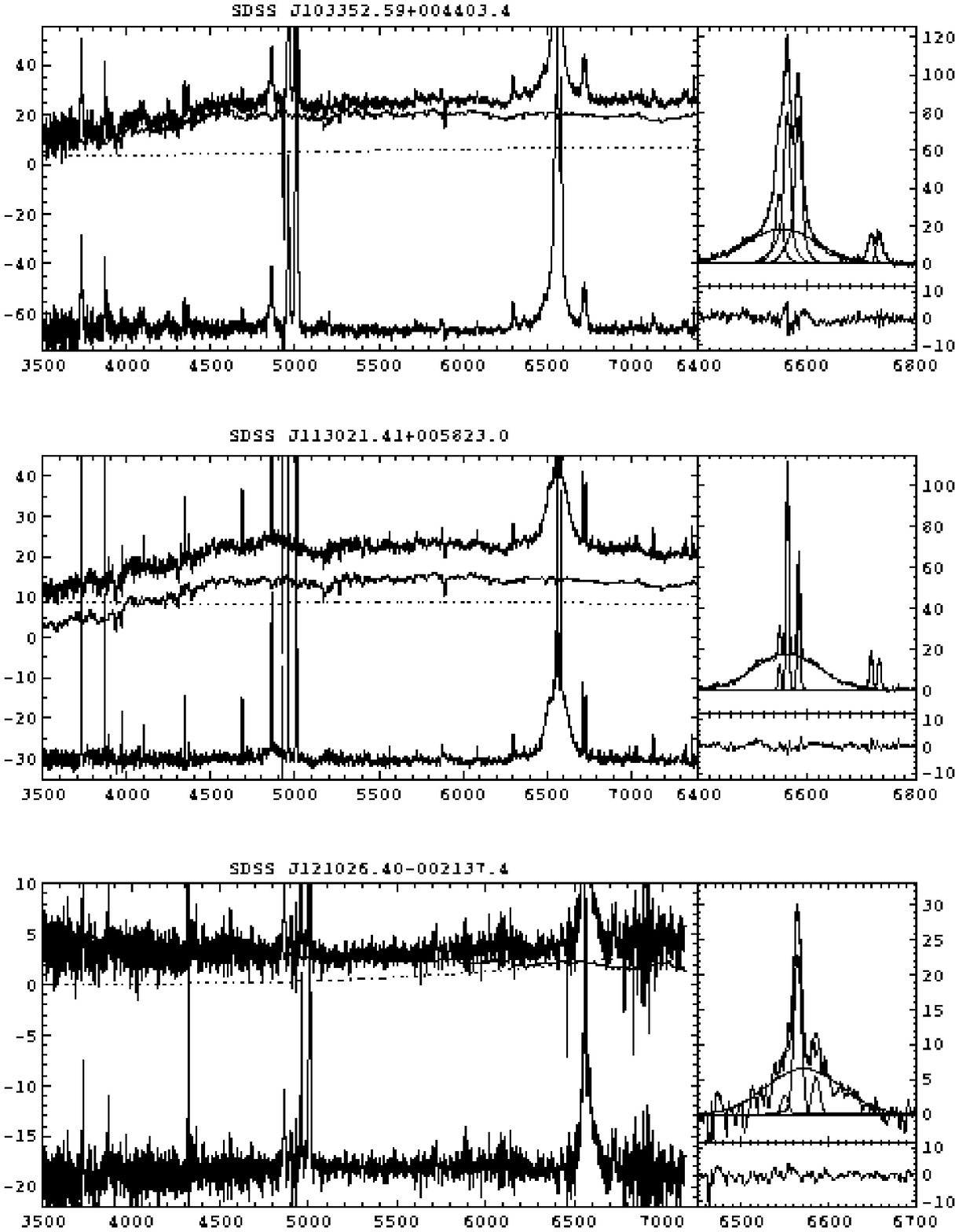} \caption{Continued...}
\end{figure}
\figurenum{1}
\begin{figure}[tbp]
\epsscale{1} \plotone{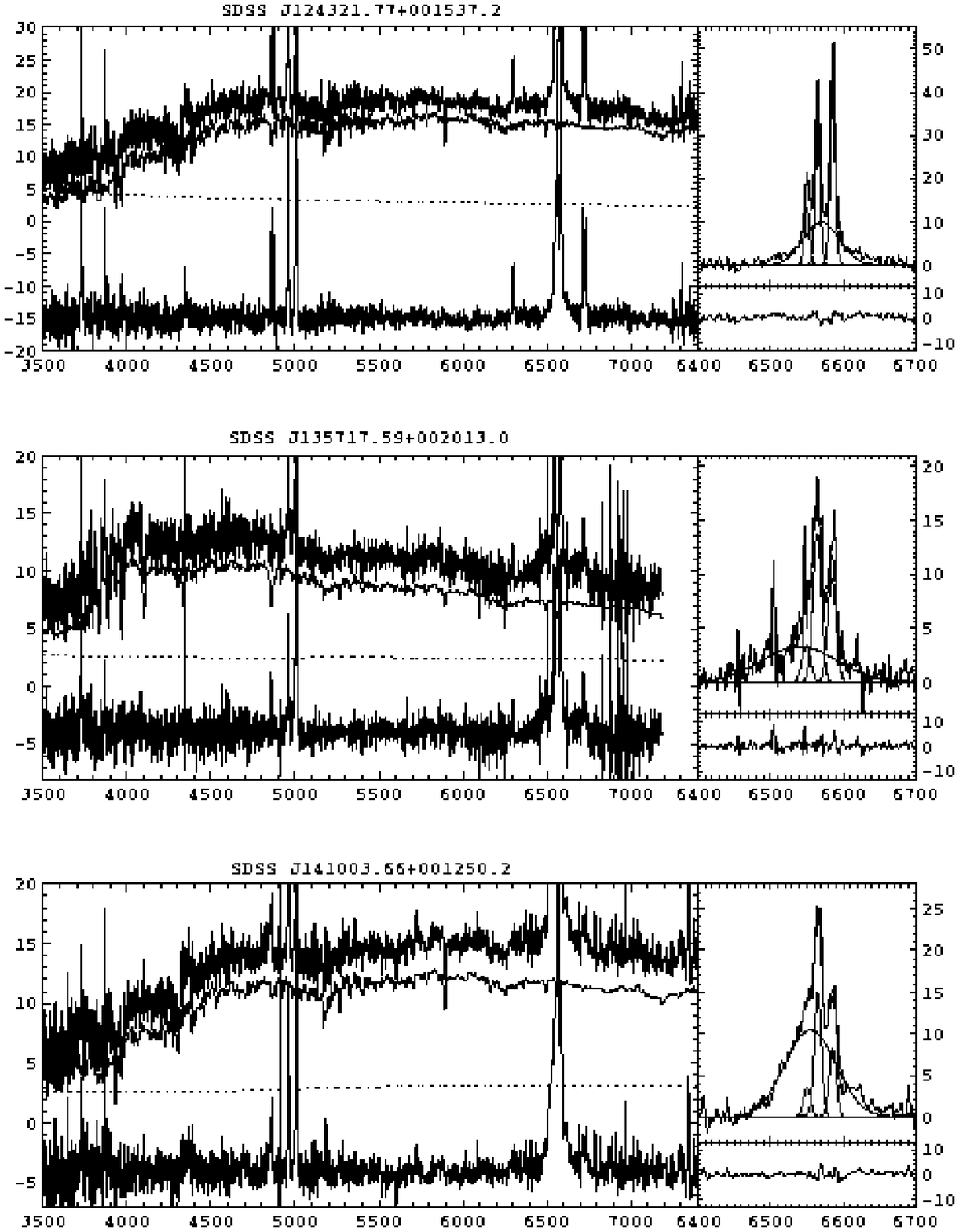} \caption{Continued...}
\end{figure}
\figurenum{1}
\begin{figure}[tbp]
\epsscale{1} \plotone{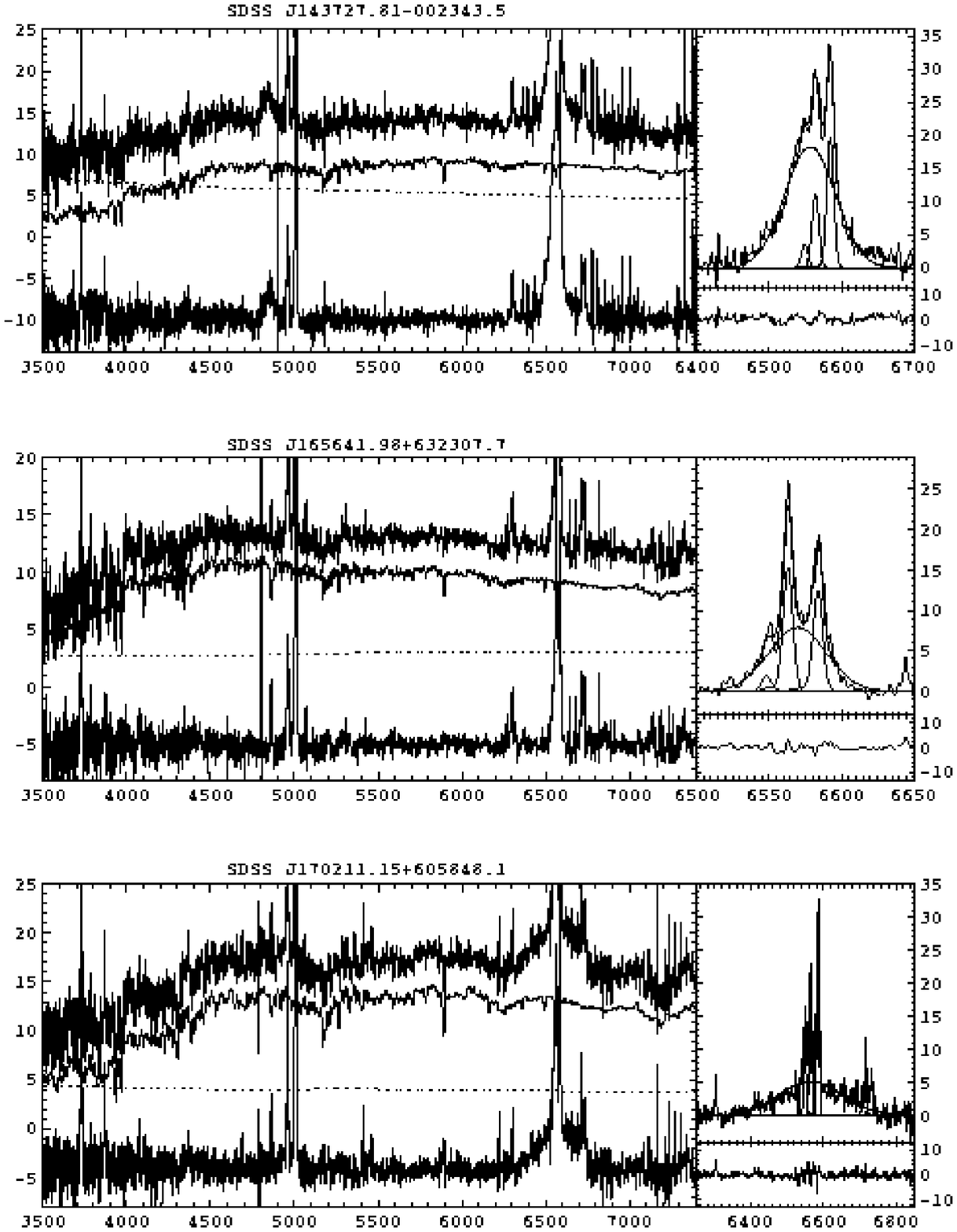} \caption{Continued...}
\end{figure}
\figurenum{1}
\begin{figure}[tbp]
\epsscale{1} \plotone{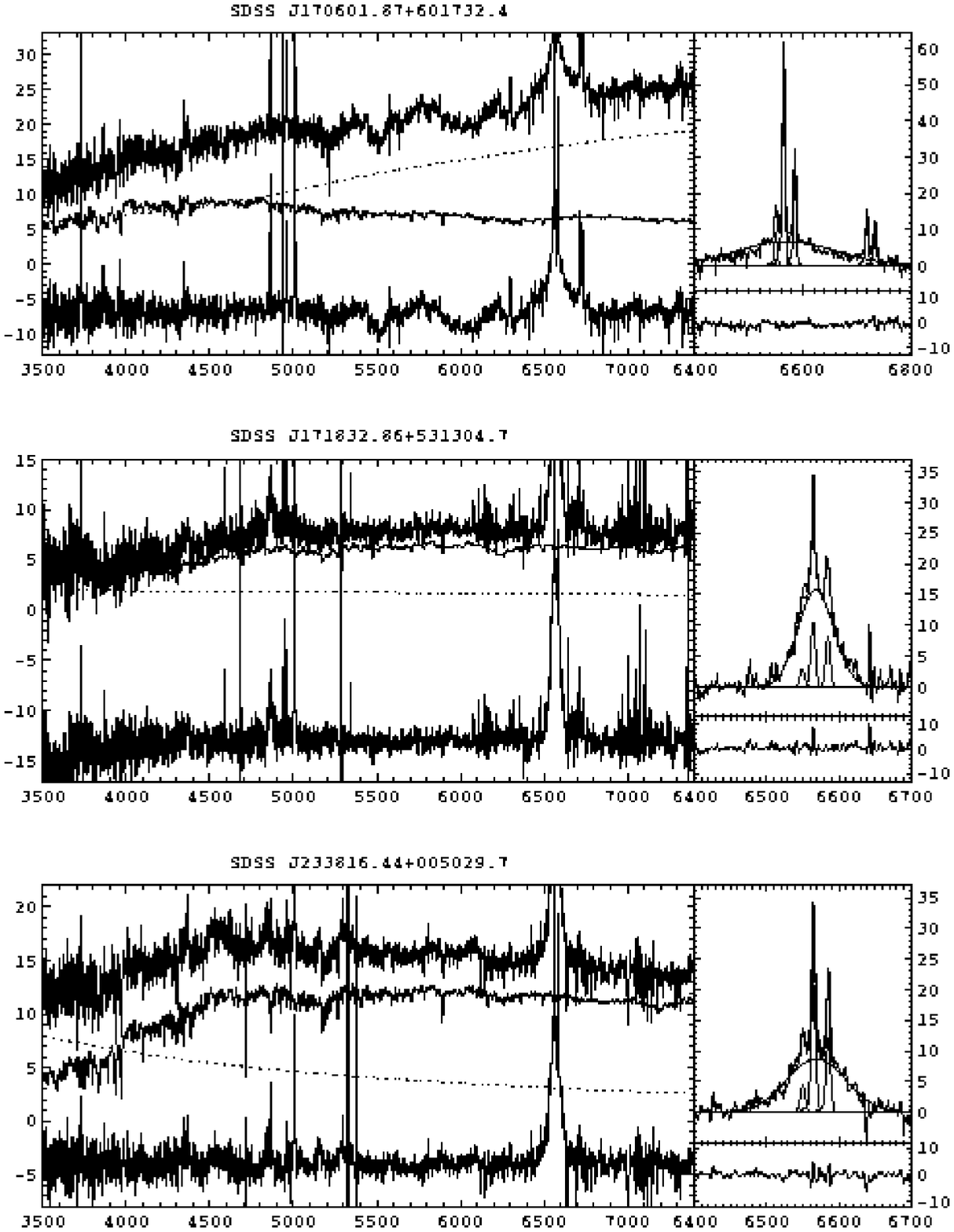} \caption{Continued...}
\end{figure}

\begin{figure}[tbp]
\figurenum{2} \epsscale{0.8} \plotone{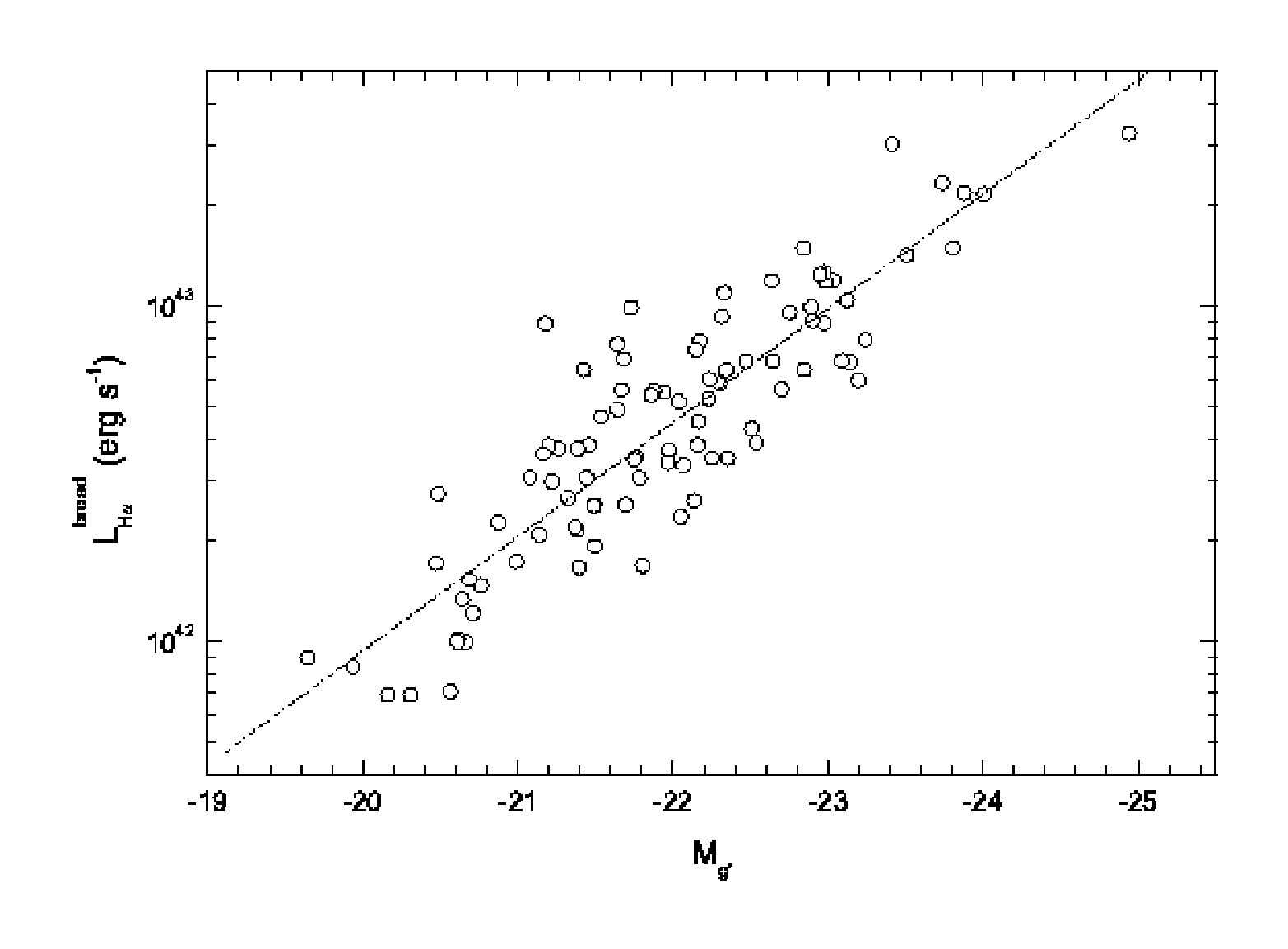} \caption{Luminosity
of the broad H$\protect\alpha$ component versus $g^{\prime}$-band
absolute magnitude for the blue AGN in the parent sample. The
dotted line shows a least square linear fitting.} \label{fig-3}
\end{figure}

\begin{figure}[tbp]
\figurenum{3} \epsscale{0.8} \plotone{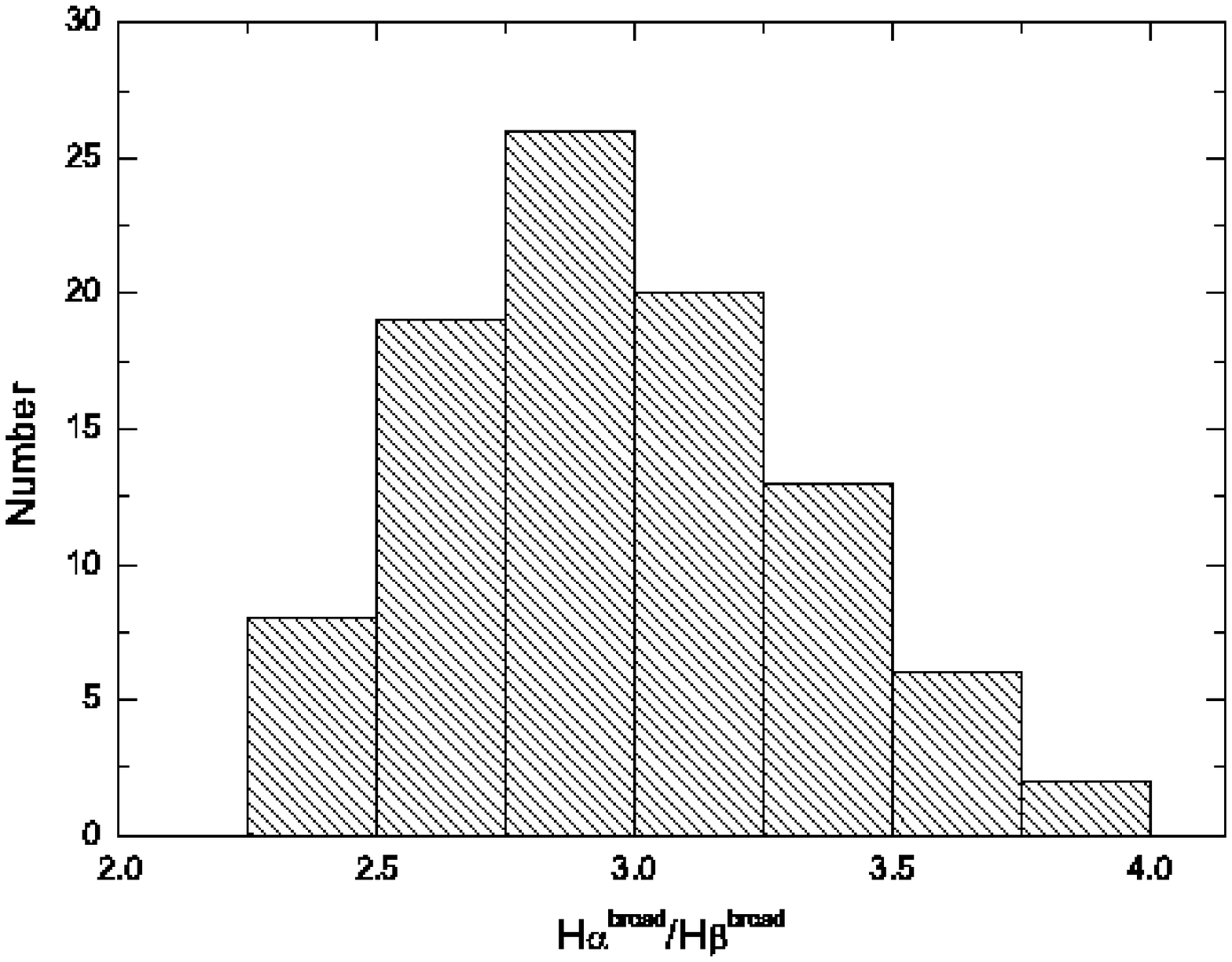} \caption{Distribution
of H$\protect\alpha $/H$\protect\beta $ ratio of their broad
components for the 94 blue AGN in the SDSS EDR. The dispersion of
H$\protect\alpha $/H$\protect\beta $ ratio is rather small.}
\label{fig-2}
\end{figure}

\begin{figure}[tbp]
\figurenum{4} \epsscale{0.8} \plotone{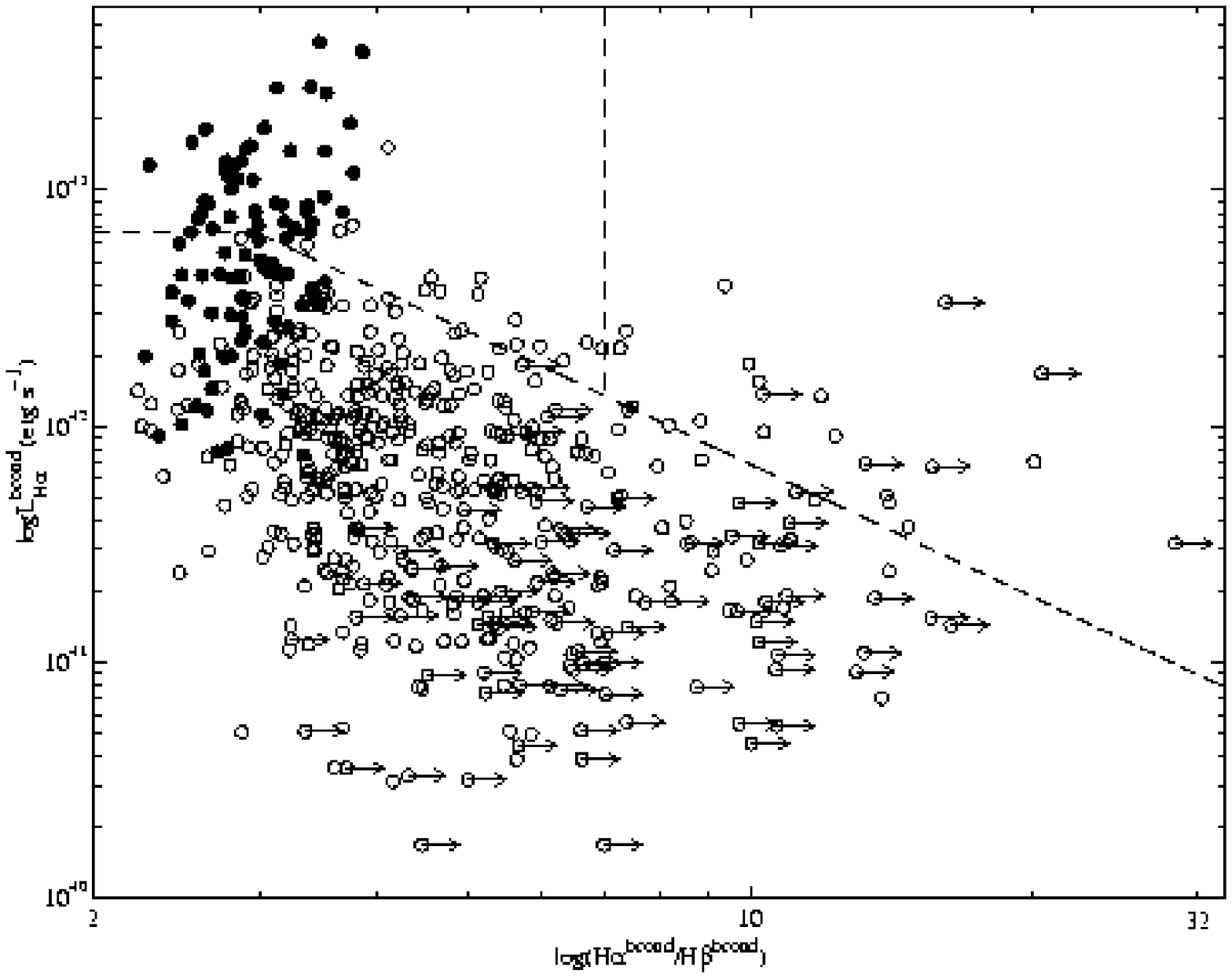} \caption{Luminosity
of the broad H$\protect\alpha $ component versus Balmer decrement
of the broad components of H$\protect\alpha $ and H$\protect\beta
$ for all the AGN in the parent sample. Those which have only
upper limit of broad H$\protect\beta $ component are tagged with a
right-pointing arrow. Blue AGN are denoted solid circles. The
inclined dash line corresponds to the estimated
M$_{g}^{nuc}=$-22$^{m}$.5 after internal extinction correction.
The vertical dash line denotes H$\protect\alpha $/H$\protect\beta
=7$. The partially obscured quasars scatter in the upper-right
region of the plot. } \label{fig-4}
\end{figure}

\begin{figure}[tbp]
\figurenum{5} \label{fig-5}\epsscale{0.8} \plotone{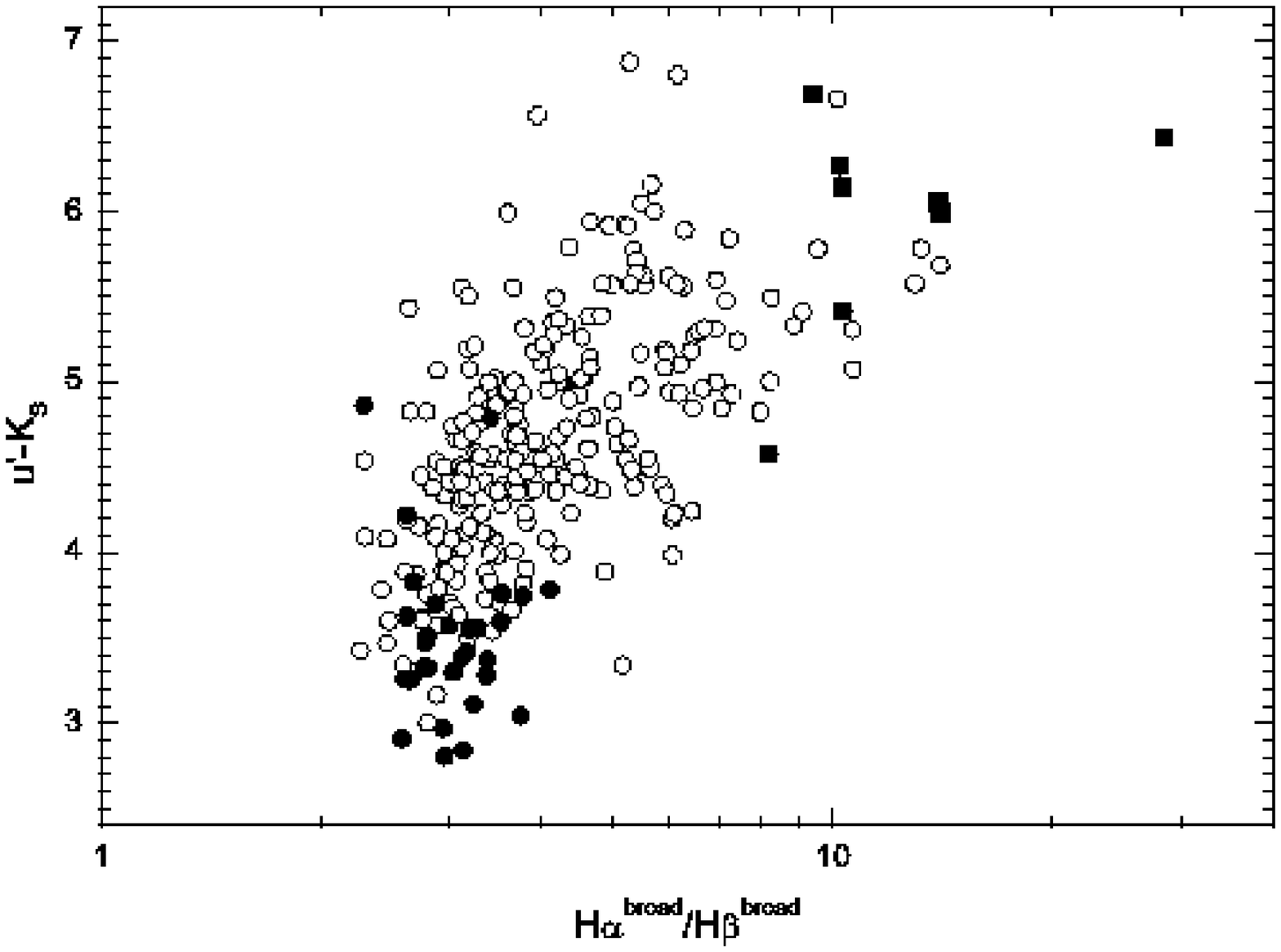}
\caption{ $u^{\prime}$-$K_{s}$ color versus the
H$\protect\alpha$/H$\protect\beta$ ratio of the broad components
for the 286 objects in the parent sample which have counterparts
in the 2MASS PSC catalog without any extended source
\textquotedblleft contamination\textquotedblright. Intermediate
type QSOs are denoted solid squares, type 1 QSOs solid circles,
and the remainder hollow circles. }
\end{figure}

\begin{figure}[tbp]
\figurenum{6} \epsscale{0.8} \plotone{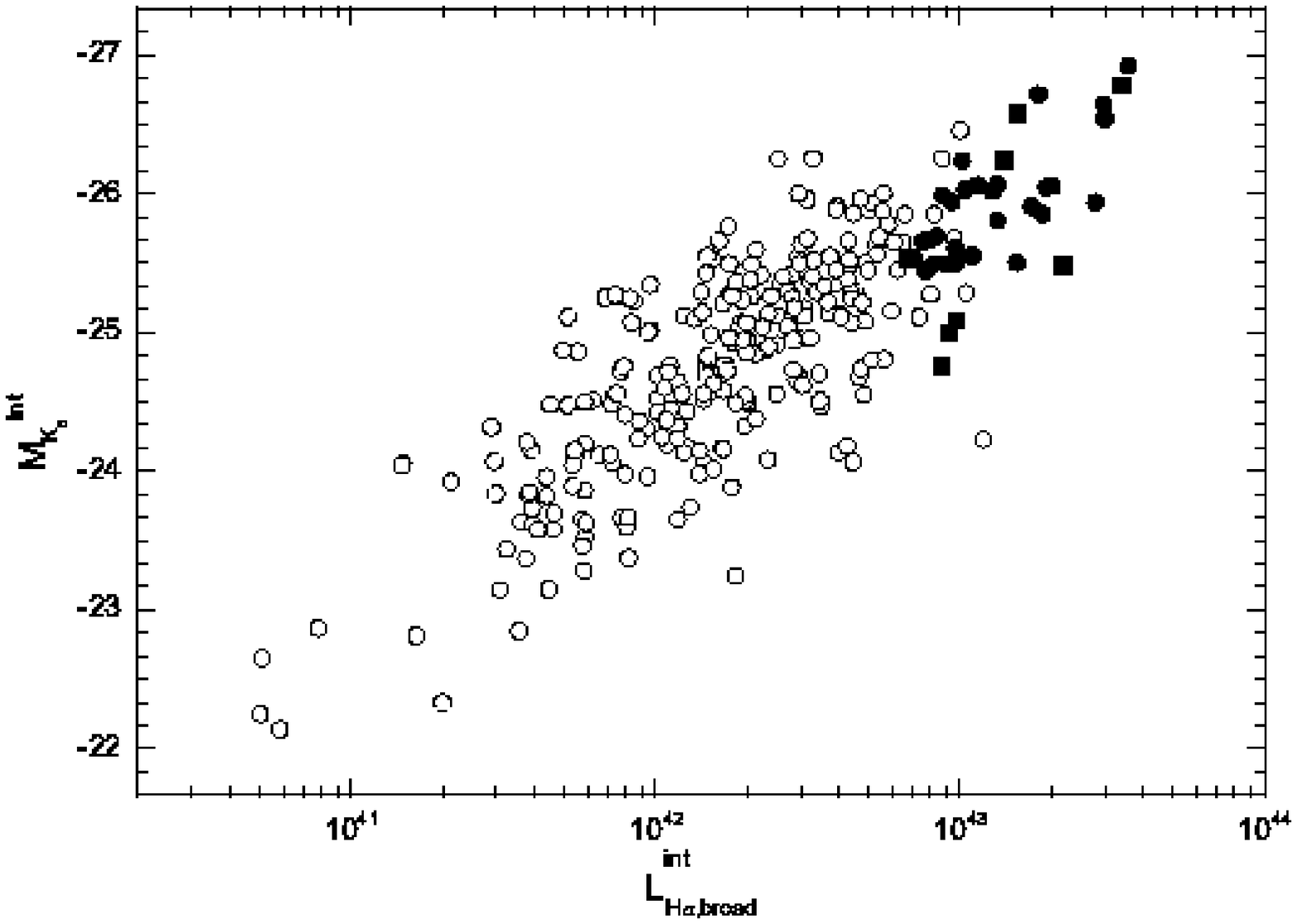} \caption{Intrinsic
absolute magnitude in the $K_{S}$ band versus the intrinsic
luminosity of the broad H$\protect\alpha $ component after
internal extinction correction for the 286 objects, denoted as in
Figure 5.} \label{fig-6}
\end{figure}

\begin{figure}[tbp]
\figurenum{7} \plotone{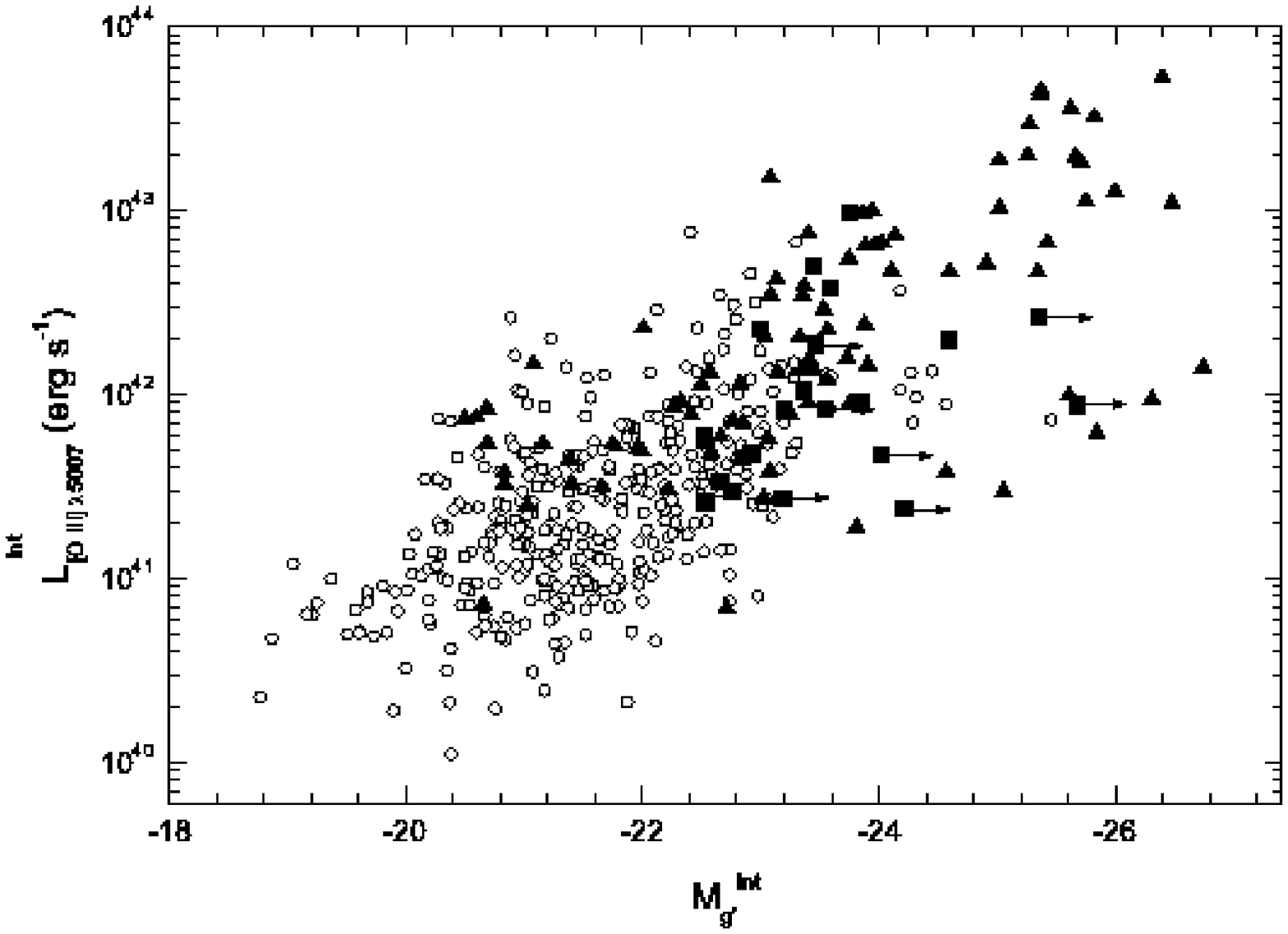} \caption{Intrinsic [O
III]$\protect\lambda $5007 luminosity versus the intrinsic
absolute magnitude in the $g^{\prime}$ band for 372 objects in the
parent sample, for which [O III] flux can be accurately measured.
The 21 intermediate type quasars (solid square) and 87 PG QSOs
(solid triangle) are also plotted for comparison. Those which have
only upper limit of broad H$\protect\beta $ component are tagged
with a right-pointing arrow. Note that the intermediate type
quasars scatter in almost the same area as PG QSOs.} \label{fig-7}
\end{figure}

\begin{figure}[tbp]
\figurenum{8} \epsscale{0.8} \plotone{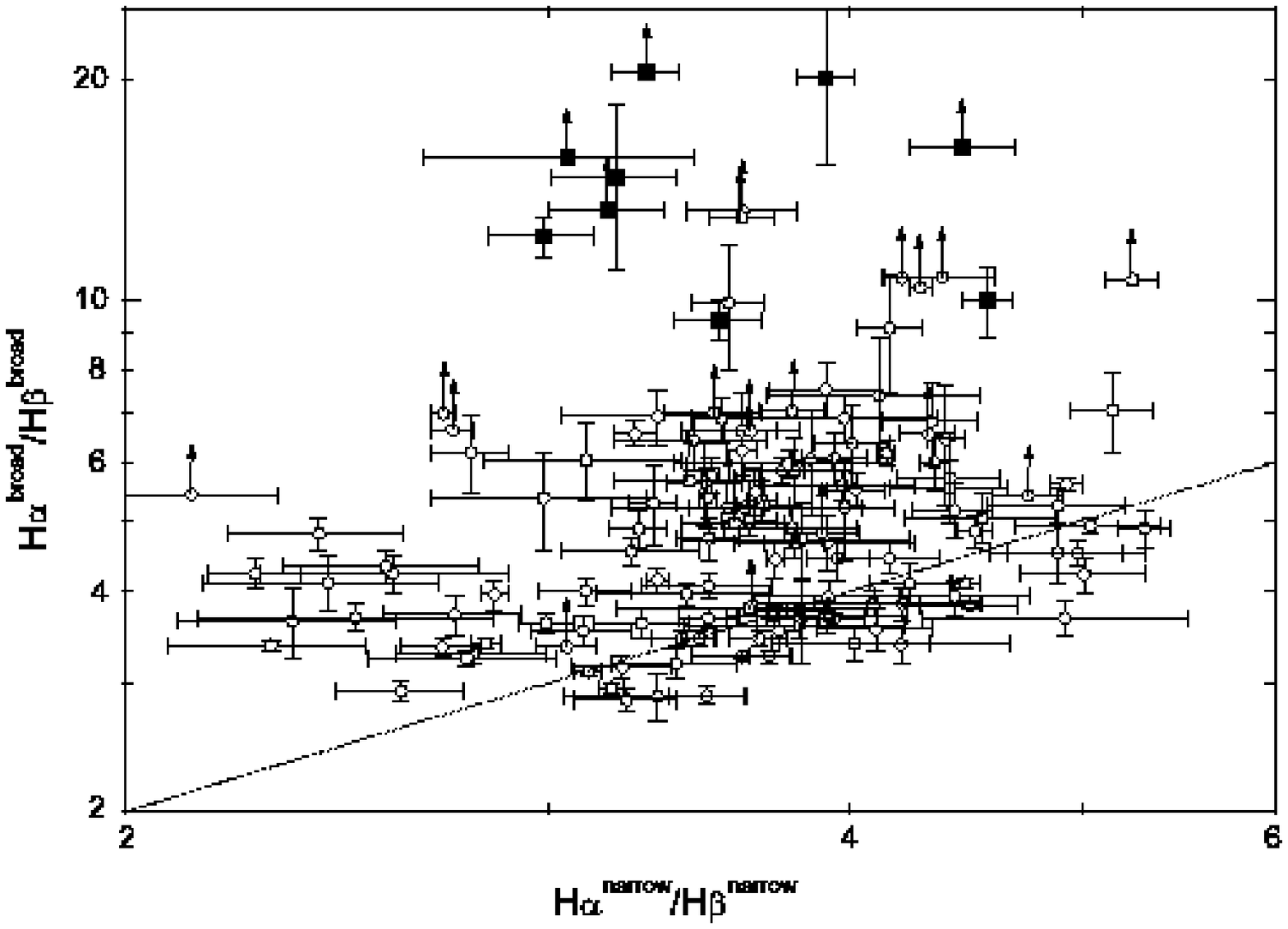} \caption{Broad
H$\protect\alpha $/H$\protect\beta $ versus narrow H$\protect
\alpha $/H$\protect\beta $ ratio of the intermediate type AGN in
the parent sample, for which the narrow H$\protect\alpha$ and
H$\protect\beta$ lines can be measured accurately. Denotements are
the same as in Figure 7.} \label{fig-8}
\end{figure}

\begin{figure}[tbp]
\figurenum{9} \epsscale{0.8} \plotone{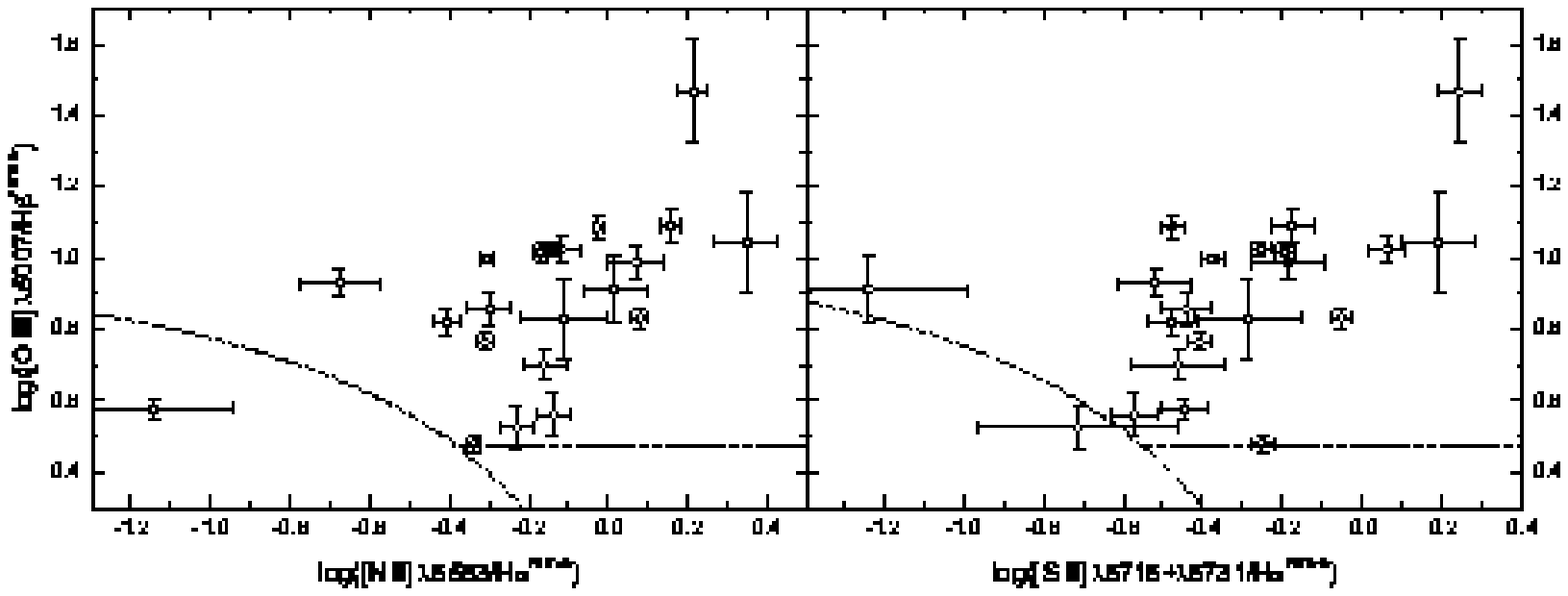} \caption{The Veilleux
\& Osterbrock (1987) diagnostic diagram of
[O III]$\protect\lambda $5007/H$\protect\beta $ versus [N II]/H$%
\protect\alpha $ (the left panel) [O III]$\protect\lambda $5007/H$\protect%
\beta $ versus [N II]/H$\protect\alpha $ (the right panel) for the
21 intermediate type quasars. The curve shows the demarcation
between HII and Seyfert galaxies defined by Kewley et al (2001)
and the horizontal straight line is $[O III]\lambda 5007/H\beta
=3$ that is often used to separate Seyferts from LINERs. Note that
all but 1 intermediate type quasars scatter in the Seyfert region.
} \label{fig-9}
\end{figure}

\begin{figure}[tbp]
\figurenum{10} \epsscale{0.65} \plotone{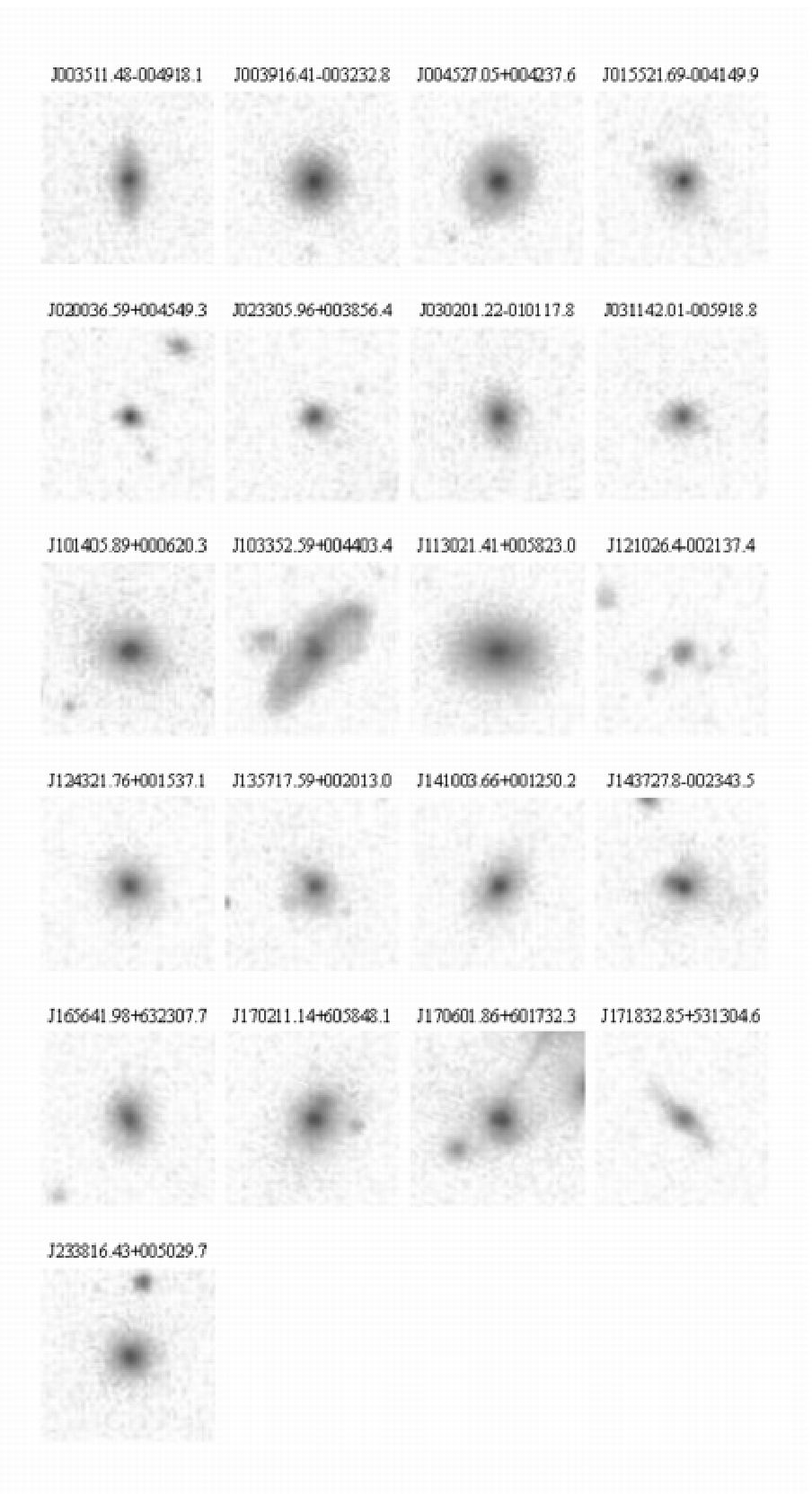} \caption{Composite
SDSS images (from $u$, $g$, $r$, $i$ and $z$ bands) of the 21
partially obscured QSO candidates. The size of each image is
21\arcsec$\times$21\arcsec. } \label{fig-10}
\end{figure}

\end{document}